\documentclass[twocolumn,english,prb,showpacs]{revtex4}
\usepackage[T1]{fontenc}
\usepackage[latin1]{inputenc}
\usepackage{amsmath}
\usepackage{graphicx}
\usepackage{amssymb}

\makeatletter

\newcommand{\noun}[1]{\textsc{#1}}
\providecommand{\tabularnewline}{\\}

\renewcommand{\vec}[1]{\mathbf{#1}}

\usepackage{babel}
\makeatother
\begin{document}

\title{Local exact exchange potentials within the all-electron FLAPW method
and a comparison with pseudopotential results}

\author{Markus Betzinger}

\email{m.betzinger@fz-juelich.de}

\affiliation{Institut f\"ur Festk\"orperforschung and Institute for Advanced
Simulation, Forschungszentrum J\"ulich and JARA, D-52425 J\"ulich,
Germany}

\author{Christoph Friedrich}

\affiliation{Institut f\"ur Festk\"orperforschung and Institute for Advanced
Simulation, Forschungszentrum J\"ulich and JARA, D-52425 J\"ulich,
Germany}

\author{Stefan Bl\"ugel}

\affiliation{Institut f\"ur Festk\"orperforschung and Institute for Advanced
Simulation, Forschungszentrum J\"ulich and JARA, D-52425 J\"ulich,
Germany}

\author{Andreas G\"orling}

\affiliation{Lehrstuhl f\"ur Theoretische Chemie, Universit\"at Erlangen-N\"urnberg,
Egerlandstr.~3, D-91058 Erlangen, Germany}

\begin{abstract}
We present a general numerical approach to construct local Kohn-Sham
potentials from orbital-dependent functionals within the all-electron
full-potential linearized augmented-plane-wave (FLAPW) method, in
which core and valence electrons are treated on an equal footing.
As a practical example, we present a treatment of the orbital-dependent
exact-exchange (EXX) energy and potential. A formulation in terms
of a mixed product basis, which is constructed from products of LAPW
basis functions, enables a solution of the optimized-effective-potential
(OEP) equation with standard numerical algebraic tools and without
shape approximations for the resulting potential. We find that the
mixed product and LAPW basis sets must be properly balanced to obtain
smooth and converged EXX potentials without spurious oscillations.
The construction and convergence of the exchange potential is analyzed
in detail for diamond. Our all-electron results for C, Si, SiC, Ge,
GaAs semiconductors as well as Ne and Ar noble-gas solids are in very
favorable agreement with plane-wave pseudopotential calculations.
This confirms the adequacy of the pseudopotential approximation in
the context of the EXX-OEP formalism and clarifies a previous contradiction
between FLAPW and pseudopotential results.
\end{abstract}


\maketitle

\section{Introduction}

Its wide applicability, accuracy, and computational efficiency have
made density-functional theory (DFT)\cite{Hohenberg-Kohn,DFT-review}
the standard method for describing the ground state of many-electron
systems. The vast majority of practical calculations employ the Kohn-Sham
(KS) formalism,\cite{Kohn-Sham} where the interacting many-electron
system is mapped onto an auxiliary noninteracting system. In this
KS system the noninteracting electrons move in a local effective potential
that is defined in such a way that the electron densities of the real
and auxiliary systems coincide. This potential is the sum of the external,
Hartree, and the exchange-correlation (xc) potentials. The latter
two contributions take into account the electron-electron interactions
including, in an indirect way, all many-body effects. The form of
the density-functional for the xc potential, which can further be
divided into an exchange and a correlation term, is unknown and must
be approximated in practice.

Fortunately, already simple approximations, like the local-density
approximation (LDA)\cite{LDA-Ceperley/Adler,LDA-VWN} or generalized
gradient approximation (GGA),\cite{GGA-PBE,GGA-PW} give reliable
results for a wide range of materials and properties. Nevertheless,
the LDA and GGA suffer from several shortcomings. First, the electrostatic
interaction of the electron with the total electron charge, described
by the Hartree potential, contains an unphysical interaction of the
electron with itself, commonly referred to as Coulomb self-interaction.
This extra term should be compensated exactly by an identical term
with opposite sign in the exchange potential, in the same manner as
in Hartree-Fock theory. However, as the LDA and GGA exchange potentials
are only approximate, this cancellation is incomplete and part of
the self-interaction remains. This error leads, in particular, to
an improper description of localized states, which appear too high
in energy and tend to delocalize. Second, the LDA and GGA functionals
do not give rise to a discontinuity of the xc potential with respect
to changes in the particle number. This discontinuity should in general
be finite (and positive), as it corresponds to the difference of the
real and the KS band gap.\cite{KS-GAP1,KS-GAP2} The latter is well
known to underestimate experimental gaps by typically 50\% or even
more. This is often called the band-gap problem of LDA and GGA. The
significance of the discontinuity for a meaningful prediction of the
fundamental band gap is discussed in Refs.~\onlinecite{KS-GAP+DIS3}
and \onlinecite{KS-GAP+DIS1}.

Functionals that depend explicitly on the electron orbitals and thus
only implicitly on the electron density form a new generation of xc
functionals.\cite{Review:O-functionals,Review-Goerling} Already the
simplest variant, the EXX functional,\cite{EXX1,EXX2,EXX3} which
treats electron exchange exactly but neglects correlation altogether,
remedies the aforementioned shortcomings of the more conventional
local and semilocal functionals: the Coulomb self-interaction term
is exactly canceled, and the local EXX potential exhibits a nonzero
discontinuity at integral particle numbers because of the orbital
dependence. 

After applications to atoms\cite{OEP-Talman-Shadwick,OEP-atoms,OEP-atoms-1}
the first implementation for periodic systems was published in 1994
by Kotani\cite{EXX-LMTO-Kotani-I} who employed the atomic sphere
approximation within the linearized muffin-tin orbital method. In
this approximation only the spherical part of the potential around
each atom is taken into account. The KS band gap turned out to be
closer to experiment than in LDA or GGA.\cite{EXX-LMTO-Kotani-I,EXX-LMTO-Kotani-II}
This indicates that the xc discontinuity and the effect of neglecting
electron correlation for these systems are roughly of the same magnitude
but of different sign and thus tend to cancel each other. Later, results
even closer to experimental values were obtained from plane-wave calculations\cite{EXX-PP-Staedele-1,EXX-PP-Staedele-2,EXX-PP-Fleszar}
employing the pseudopotential (PP) approximation, which allows accurate
treatment of the warped shape of the EXX potential except for the
regions close to the atomic nuclei where the potential is smoothed.

However, when the first all-electron (AE), full-potential results
were reported,\cite{Sharma} they deviated substantially from the
PP values and were also in considerably worse agreement with experiment.
In Ref.~\onlinecite{Sharma}, Sharma \textit{et al.}, who implemented
the EXX functional within the FLAPW method, argue that the success
of EXX in the earlier PP calculations is only an artifact of the neglect
of the core-valence exchange interaction. They conclude that treatment
of core and valence electrons on the same footing is imperative for
a proper EXX calculation. This work started a controversy about the
adequacy of the PP approximation in the EXX formalism. Recently, Engel\cite{Engel}
showed that, on the contrary, AE and PP results for lithium and diamond
differ only marginally. The AE calculations were performed with a
plane-wave basis by pushing the PP and plane-wave cutoffs to the AE
limit. Also recently, Makmal \textit{et al.} reported a similarly
good agreement between AE and PP calculations for the diatomic molecules
BeO and CO using real-space grid approaches.\cite{Makmal}

A final comparison between PP results and results obtained from a
genuine AE approach for periodic systems, such as the FLAPW method,
is still missing, though. With this work we want to fill this gap.
We present an alternative implementation of the EXX approach within
the FLAPW method, which uses a numerical approach different from the
one reported in Ref.~\onlinecite{Sharma}. It employs a specifically
designed basis, the mixed product basis, in which the optimized-effective-potential
(OEP) equation for the EXX potential is solved. The mixed product
functions form an all-electron basis for the products of single-particle
wave functions occurring in the OEP equation. Within this approach,
both spherical and nonspherical as well as warped interstitial contributions
to the EXX potential are fully taken into account. For the example
of diamond, we discuss in detail the convergence of the local EXX
potential and the resulting KS band gaps with respect to the quality
of the mixed product and LAPW basis sets. We demonstrate that a smooth
potential and a direct KS band gap very close to the result by Engel\cite{Engel}
are obtained if the two basis sets are properly balanced. A similar
behavior has been reported for Gaussian and plane-wave basis sets.\cite{Balance-Gaussian,Hesselmann,Comparison-OEP-KS}
To validate our findings, we also report results for Si, SiC, Ge,
GaAs, and crystalline Ne and Ar. For all materials, we find a very
good agreement between our AE and previously published plane-wave
PP values. We conclude that the large discrepancies found in Ref.~\onlinecite{Sharma}
cannot be attributed to the core-valence interaction.

The paper is organized as follows. Sections \ref{sec: Theory} and
\ref{sec: FLAPW-Method} give brief introductions into the theory
and the FLAPW method. Our implementation of the EXX functional within
the FLAPW program \noun{Fleur}\cite{Fleur} is described in Section
\ref{sec: Implementation}. Section \ref{sec: Results-&-Discussion}
discusses the convergence of the effective potential for the example
of diamond and compares AE EXX KS eigenvalue differences commonly
interpreted as transition energies for C, Si, SiC, Ge, GaAs, crystalline
Ne, and Ar with theoretical plane-wave and experimental values from
the literature. Finally, we draw our conclusions in Section \ref{sec: Conclusions}.

\section{Theory\label{sec: Theory}}

The KS formalism\cite{Kohn-Sham} of DFT\cite{Hohenberg-Kohn} relies
on an auxiliary system of noninteracting electrons, which move in
the spin-dependent local effective potential \begin{equation}
V_{\mathrm{eff}}^{\sigma}(\vec{r})=V_{\mathrm{ext}}(\vec{r})+V_{\mathrm{H}}(\vec{r})+V_{\mathrm{xc}}^{\sigma}(\vec{r})\end{equation}
with the external, Hartree, and xc potential, respectively. The latter
is defined in such a way that the electron spin densities coincide
with those of the real interacting system. It is given by the functional
derivative of the xc energy functional $E_{\mathrm{xc}}[n^{\uparrow},n^{\downarrow}]$
with respect to the electron spin density $n^{\sigma}(\vec{r})$ $(\sigma=\uparrow,\downarrow)$,\begin{equation}
V_{\mathrm{xc}}^{\sigma}(\vec{r})=\frac{\delta E_{\mathrm{xc}}}{\delta n^{\sigma}(\vec{r})}\,.\label{eq: definition V_xc}\end{equation}
The orbitals describing the electrons in the auxiliary system obey
the KS equations\begin{equation}
\left[-\frac{1}{2}\nabla^{2}+V_{\mathrm{eff}}^{\sigma}(\vec{r})\right]\varphi_{n\vec{k}}^{\sigma}(\vec{r})=\epsilon_{n\vec{k}}^{\sigma}\varphi_{n\vec{k}}^{\sigma}(\vec{r})\,,\label{eq: KS Hamiltonian}\end{equation}
where $\varphi_{n\vec{k}}^{\sigma}$ denotes the KS orbital of spin
$\sigma$, band index $n$ and Bloch vector $\vec{k}$. Hartree atomic
units are used except where explicitly noted. The electron spin density
is given by a sum over the occupied states \begin{eqnarray}
n^{\sigma}(\vec{r}) & = & \sum_{\vec{k}}\sum_{n}^{\mathrm{occ.}}|\varphi_{n\vec{k}}^{\sigma}(\vec{r})|^{2}\,.\label{eq: density}\end{eqnarray}
By a summation over Bloch vectors we mean an integration over the
Brillouin zone, which is sampled by a finite set of mesh points.

For the conventional LDA and GGA functionals, the xc energy functional
depends locally on the spin densities and, in the case of GGA, on
the their gradients, and the functional derivative in Eq.~\eqref{eq: definition V_xc}
translates to a derivative of a function and is evaluated in a straightforward
way. However, for orbital-dependent functionals, $E_{\mathrm{xc}}$
depends only indirectly on the electron spin densities: the KS orbitals,
which define $E_{\mathrm{xc}}\left[\varphi^{\uparrow},\varphi^{\downarrow}\right]$,
are functionals of the effective potential $V_{\mathrm{eff}}^{\sigma}(\vec{r})$
through Eq.~\eqref{eq: KS Hamiltonian}, and $V_{\mathrm{eff}}^{\sigma}(\vec{r})$
is a functional of $n^{\sigma}$. Therefore, one must apply the chain
rule to calculate the functional derivative in Eq.~\eqref{eq: definition V_xc}\begin{eqnarray}
\lefteqn{V_{\mathrm{xc}}^{\mathrm{\sigma}}(\vec{r})}\label{eq: V_xc chain rule}\\
 & = & \sum_{n,\vec{k}}\iint\left[\frac{\delta E_{\mathrm{xc}}}{\delta\varphi_{n\vec{k}}^{\sigma}(\vec{r}')}\frac{\delta\varphi_{n\vec{k}}^{\sigma}(\vec{r}')}{\delta V_{\mathrm{eff}}^{\sigma}(\vec{r}'')}+\mathrm{c.c.}\right]\frac{\delta V_{\mathrm{eff}}^{\sigma}(\vec{r}'')}{\delta n^{\sigma}(\vec{r})}d^{3}r'\, d^{3}r''\nonumber \end{eqnarray}
where the sum runs over all KS states present in $E_{\mathrm{xc}}$.
Then, multiplication with the single-particle spin-density response
function\begin{equation}
\chi_{\mathrm{s}}^{\sigma}(\vec{r},\vec{r}')=\frac{\delta n^{\sigma}(\vec{r})}{\delta V_{\mathrm{eff}}^{\sigma}(\vec{r}')}\,,\label{eq: response function}\end{equation}
integration, and use of $\chi_{\mathrm{s}}^{\sigma}(\vec{r},\vec{r}')=\chi_{\mathrm{s}}^{\sigma}(\vec{r}',\vec{r})$
yields an integral equation for the xc potential\begin{eqnarray}
\lefteqn{\int\chi_{\mathrm{s}}^{\sigma}(\vec{r},\vec{r}')V_{\mathrm{xc}}^{\sigma}(\vec{r}')d^{3}r'}\nonumber \\
 & = & \sum_{\vec{k}}\sum_{n}\int\left[\frac{\delta E_{\mathrm{xc}}}{\delta\varphi_{n\vec{k}}^{\sigma}(\vec{r}')}\frac{\delta\varphi_{n\vec{k}}^{\sigma}(\vec{r}')}{\delta V_{\mathrm{eff}}^{\sigma}(\vec{r})}+\mathrm{c.c.}\right]d^{3}r'\,.\label{eq: general OEP equation}\end{eqnarray}
In this work we employ, as a practical example, the orbital-dependent
EXX functional

\begin{eqnarray}
E_{\mathrm{x}} & = & -\frac{1}{2}\sum_{\sigma}\sum_{\vec{k},\vec{q}}\sum_{n,n'}^{\mathrm{occ.}}\iint d^{3}r\, d^{3}r'\nonumber \\
 &  & \times\frac{\varphi_{n\vec{k}}^{\sigma*}(\vec{r})\varphi_{n'\vec{q}}^{\sigma}(\vec{r})\varphi_{n'\vec{q}}^{\sigma*}(\vec{r}')\varphi_{n\vec{k}}^{\sigma}(\vec{r}')}{|\vec{r}-\vec{r}'|}\label{eq: EXX functional}\end{eqnarray}
whose functional derivative with respect to the KS wave functions
is given by the well-known Hartree-Fock expression\begin{eqnarray}
\frac{\delta E_{\mathrm{x}}}{\delta\varphi_{n\vec{k}}^{\sigma}(\vec{r}')} & = & \int\varphi_{n\vec{k}}^{\sigma*}(\vec{r}'')V_{\mathrm{x},\mathrm{NL}}^{\sigma}(\vec{r}'',\vec{r}')d^{3}r''\label{eq: HF term}\end{eqnarray}
with\begin{equation}
V_{\mathrm{x,NL}}^{\sigma}(\vec{r}'',\vec{r}')=-\sum_{\vec{q}}\sum_{n'}^{\mathrm{occ.}}\frac{\varphi_{n'\vec{q}}^{\sigma}(\vec{r}'')\varphi_{n'\vec{q}}^{\sigma*}(\vec{r}')}{|\vec{r}'-\vec{r}''|}\,.\label{eq: nonlocal HF kernel}\end{equation}
First-order perturbation theory yields the wave-function response
\begin{equation}
\frac{\delta\varphi_{n\vec{k}}^{\sigma}(\vec{r})}{\delta V_{\mathrm{eff}}^{\sigma}(\vec{r}')}=\sum_{n'(\ne n)}\frac{\varphi_{n'\vec{k}}^{\sigma*}(\vec{r}')\varphi_{n\vec{k}}^{\sigma}(\vec{r}')}{\epsilon_{n\vec{k}}^{\sigma}-\epsilon_{n'\vec{k}}^{\sigma}}\varphi_{n'\vec{k}}^{\sigma}(\vec{r})\label{eq: wave-function response}\end{equation}
and together with Eq.~\eqref{eq: density} the spin-density response
function \begin{equation}
\chi_{s}^{\sigma}(\vec{r},\vec{r}')=2\sum_{\vec{k}}\sum_{n}^{\mathrm{occ.}}\sum_{n'}^{\mathrm{unocc.}}\frac{\varphi_{n\vec{k}}^{\sigma*}(\vec{r})\varphi_{n'\vec{k}}^{\sigma}(\vec{r})\varphi_{n'\vec{k}}^{\sigma*}(\vec{r}')\varphi_{n\vec{k}}^{\sigma}(\vec{r}')}{\epsilon_{n\vec{k}}^{\sigma}-\epsilon_{n'\vec{k}}^{\sigma}}\label{eq: response explicit}\end{equation}
where time-reversal symmetry has been used. Using Eqs.~\eqref{eq: HF term},
\eqref{eq: wave-function response}, and \eqref{eq: response explicit}
the integral equation {[}Eq.~\eqref{eq: general OEP equation}] turns
into\begin{eqnarray}
\int\chi_{\mathrm{s}}^{\sigma}(\vec{r},\vec{r}')V_{\mathrm{x}}^{\sigma}(\vec{r}')d^{3}r' & = & t^{\sigma}(\vec{r})\label{eq: x OEP equation}\end{eqnarray}
with\begin{align}
t^{\sigma}(\vec{r}) & =\frac{\delta E_{\mathrm{x}}}{\delta V_{\mathrm{eff}}^{\sigma}(\vec{r})}\nonumber \\
 & =2\sum_{\vec{k}}\sum_{n}^{\mathrm{occ.}}\sum_{n'}^{\mathrm{unocc.}}\left[\langle\varphi_{n\vec{k}}^{\sigma}|V_{\mathrm{x,\mathrm{NL}}}^{\sigma}|\varphi_{n'\vec{k}}^{\sigma}\rangle\frac{\varphi_{n'\vec{k}}^{\sigma*}(\vec{r})\varphi_{n\vec{k}}^{\sigma}(\vec{r})}{\epsilon_{n\vec{k}}^{\sigma}-\epsilon_{n'\vec{k}}^{\sigma}}\right]\label{eq: rhs t explicit}\end{align}
and\begin{eqnarray}
\lefteqn{\langle\varphi_{n\vec{k}}^{\sigma}|V_{\mathrm{x,\mathrm{NL}}}^{\sigma}|\varphi_{n'\vec{k}}^{\sigma}\rangle}\label{eq: matrix elements of V_x_NL}\\
 & = & \iint\varphi_{n\vec{k}}^{\sigma*}(\vec{r})V_{\mathrm{x,NL}}^{\sigma}(\vec{r},\vec{r}')\varphi_{n'\vec{k}}^{\sigma}(\vec{r}')d^{3}r\, d^{3}r'\,.\nonumber \end{eqnarray}
In this form the integral equation is called OEP equation and goes
back to Sharp and Horton,\cite{OEP/OPM} who derived Eq.~\eqref{eq: x OEP equation}
as a result of a variational minimization of the Hartree-Fock total
energy under the additional constraint that the orbitals experience
a local rather than a nonlocal potential. Sahni \emph{et al.}\cite{OEP<=>EXX}
finally realized that the OEP approach is equivalent to the construction
of a local EXX potential within the KS formalism.

\section{FLAPW Method\label{sec: FLAPW-Method}}

The LAPW basis\cite{FLAPW1,FLAPW2,FLAPW3} is constructed from piecewise
defined functions to deal, at the same time, with the atomic-like
potential close to the nuclei and the smooth potential in the region
far away from the nuclei. For this purpose, space is partitioned into
nonoverlapping atom-centered muffin-tin (MT) spheres and the remaining
interstitial region (IR), where the smoothness of the potential allows
to employ plane waves as basis functions. At the MT sphere boundaries,
these plane waves are matched in value and first radial derivative
to linear combinations of spin-dependent MT solutions $u_{l0}^{a\sigma}(r)Y_{lm}(\hat{\vec{r}})$
of the radial scalar-relativistic Dirac equation and their energy
derivatives $u_{l1}^{a\sigma}(r)Y_{lm}(\hat{\vec{r}})$ using the
spherical average of the effective potential and predefined energy
parameters that lie in the energy range of the occupied states. Here,
$Y_{lm}(\hat{\vec{r}})$ denote the spherical harmonics, $\vec{r}$
is measured from the MT center of atom $a$ and $\hat{\vec{r}}=\vec{r}/r$
is a unit vector. This gives the LAPW basis functions\begin{widetext}\begin{equation}
\phi_{\vec{k}\vec{G}}^{\sigma}(\vec{r})=\left\{ \begin{array}{cl}
\frac{1}{\sqrt{\Omega}}\exp\left[i(\vec{k}+\vec{G})\cdot\vec{r}\right] & \mathrm{if\,}\vec{r}\in\mathrm{IR}\\
\sum_{l=0}^{l_{\mathrm{max}}}\sum_{m=-l}^{l}\sum_{p=0}^{1}A_{lmp}^{\sigma}(\vec{k},\vec{G})u_{lp}^{a\sigma}(|\vec{r}-\vec{R}_{a}|)Y_{lm}(\widehat{\vec{r}-\vec{R}_{a}}) & \mathrm{if\,}\vec{r}\in\mathrm{MT}(a)\end{array}\right.\label{eq: APWs}\end{equation}
\end{widetext}for the valence electrons with the unit-cell volume
$\Omega$ and reciprocal lattice vectors $\vec{G}$. For a practical
calculation cutoff values for the reciprocal lattice vectors $|\vec{k}+\vec{G}|\le G_{\mathrm{max}}$
and the angular momentum $l\le l_{\mathrm{max}}$ are employed. The
core states are obtained by solving the fully relativistic Dirac equation
with the spherical average of the effective potential\@.

The basis functions, defined in Eq.~\eqref{eq: APWs}, can represent
only those wave functions accurately whose energies are sufficiently
close to the energy parameters, which are usually located in the valence-band
region. For a precise description of semicore and high-lying unoccupied
states that are far away from the energy parameters the basis must
be augmented and local orbitals (lo)\cite{Local_Orbitals1,Local_Orbitals2,Local_Orbitals3}
are currently the best developed technique. Let us assume that we
want to improve the basis for states with an angular momentum $l$
around an energy $\epsilon^{\mathrm{lo}}$. Then we construct an additional
radial function $u_{l}^{a\sigma}(r,\epsilon^{\mathrm{lo}})$ from
the radial scalar-relativistic Dirac equation with the energy parameter
$\epsilon^{\mathrm{lo}}$ and form a linear combination $u_{lp}^{a\sigma}(r)Y_{lm}(\hat{\vec{r}})$
($p\ge2$, the index $p$ is a label numbering the basis functions
for a given $l$, $a$, and $\sigma$) from $u_{l}^{a\sigma}(r,\epsilon^{\mathrm{lo}})$
and the radial functions $u_{l0}^{a\sigma}(r)$ and $u_{l1}^{a\sigma}(r)$,
already defined above, such that $u_{lp}^{a\sigma}(r)$ is normalized
and its value and radial derivative vanish at the MT boundary. In
this way, the local orbital $u_{lp}^{a\sigma}(r)Y_{lm}(\hat{\vec{r}})$
is completely confined to the MT sphere and need not be matched to
a plane wave outside. For semicore states, which are nearly dispersion-less,
the energy parameter $\epsilon^{\mathrm{lo}}$ is fixed at the semicore
energy level. For the unoccupied states we use energy parameters chosen
such that the solutions of the radial scalar-relativistic Dirac equation
fulfill\begin{equation}
\left.\frac{d}{dr}\mathrm{ln}[u_{l}^{a\sigma}(r,\epsilon^{\mathrm{lo}})]\right|_{r=S}=-(l+1)\end{equation}
at the MT sphere boundary $r=S$, following a procedure proposed in
Ref.~\onlinecite{Local_Orbitals_Andersen}. This condition yields
for each $l$ quantum number a series of orthogonal solutions of increasing
energies. We use the resulting local orbitals to converge the LAPW
basis in a systematic way.

\section{Implementation\label{sec: Implementation}}

To solve the integral equation {[}Eq.~\eqref{eq: x OEP equation}],
we introduce a basis $\{ M_{I}(\vec{r})\}$ that reformulates the
equation as a linear-algebra problem \begin{equation}
\sum_{J}\chi_{\mathrm{s},IJ}^{\sigma}V_{\mathrm{x},J}^{\sigma}=t_{I}^{\sigma}\,,\label{eq: algebraic OEP}\end{equation}
which can be solved for the exchange potential $V_{\mathrm{x}}^{\sigma}$
by matrix inversion of $\chi_{\mathrm{s}}^{\sigma}$ applying standard
numerical techniques. As all quantities appearing in Eq.~\eqref{eq: x OEP equation}
are defined in terms of wave-function products, the basis should be
constructed foremost of products of LAPW basis functions. In recent
publications we have already used such a mixed product basis (MPB),
which was first proposed by Kotani and van Schilfgaarde,\cite{Kotani-MixedBasis}
to implement hybrid functionals\cite{PBE0-NonLocalExactExchangePotential}
and the \textit{GW} approximation\cite{GW-MixedBasis} as well as
calculate EELS spectra.\cite{CoulombMatrix-MixedBasis} However, we
will introduce a slightly modified version for the present purpose:
(1) since the potential is strictly periodic, the MPB may be restricted
to $\vec{k}=\vec{0}$, (2) we add the atomic exact exchange potential
as a basis function, and (3) we make the functions $M_{I}(\vec{r})$
continuous over the whole space.

The construction of the MPB and the implementation of the spin-density
response function $\chi_{s,IJ}^{\sigma}$ and $t_{I}^{\sigma}$ are
described in Sec.~\ref{sub: Auxiliary-basis-set} and \ref{sub: Single-particle-response},
respectively. Numerical tests of the implementation are shown in Sec.~\ref{sub: Tests}.

\subsection{Mixed product basis\label{sub: Auxiliary-basis-set}}

The MPB consists of plane waves in the IR and MT functions $M_{LP}^{a}(r)Y_{LM}(\hat{\vec{r}})$
in the spheres that derive from products of the functions $u_{lp}^{a\sigma}(r)Y_{lm}(\hat{\vec{r}})$.
As in the LAPW basis, cutoff values $G'_{\mathrm{max}}$ for the interstitial
plane waves and $L_{\mathrm{max}}$ for the angular momentum quantum
numbers $L$ are employed. For mathematical details of the construction
of the MPB we refer the reader to our previous publications Refs.~\onlinecite{CoulombMatrix-MixedBasis,GW-MixedBasis,PBE0-NonLocalExactExchangePotential}.
Here, we lay emphasis on the modifications for the present implementation
of the EXX-OEP method.

From EXX-OEP calculations of atoms it is known that the local exchange
potential shows pronounced humps which reflect the atomic shell structure.\cite{OEP-atoms}
As the electron orbitals contract spatially for atoms with larger
atomic numbers, these humps move closer and closer to the atomic nucleus.
Near the nucleus, the exchange potential of a periodic crystal resembles
that of the corresponding atom, because the long-range exchange interactions
with electrons on neighboring atoms contribute only a slowly varying
potential there. Therefore, we augment the MPB with the spherical
exchange potential from an atomic EXX-OEP calculation performed with
the relativistic atomic structure program RELKS.\cite{RELKS1,RELKS2,RELKS3}
It is added to the set of spherical MT functions $(L=0)$. The rest
of the basis must then only describe the difference between the atomic
and the crystal exchange potential. In this way all nonlocal exchange
contributions are fully taken into account.

To avoid discontinuities of the resulting potential at the MT sphere
boundaries, we form linear combinations of the MT functions and interstitial
plane waves that are continuous in value and first derivative there.
In analogy to the construction of the LAPW basis (s.~Sec.~\ref{sec: FLAPW-Method}),
two radial functions per $lm$ channel are used to augment the interstitial
plane waves in the MT spheres, while the remaining functions are combined
to form local orbitals. We note that there are usually far more than
two radial functions per $lm$ channel in the MPB. We also note that
such a construction was not needed in our earlier implementations.

\subsection{Spin-density response function and $t_{I}^{\sigma}(\vec{r})$\label{sub: Single-particle-response}}

In the MPB the spin-density response function in Eq.~\eqref{eq: response explicit}
and $t_{I}^{\sigma}(\vec{r})$ in Eq.~\eqref{eq: rhs t} become\begin{equation}
\chi_{\mathrm{s},IJ}^{\sigma}=2\sum_{\vec{k}}\sum_{n}^{\mathrm{occ.}}\sum_{n'}^{\mathrm{unocc.}}\frac{\langle M_{I}\varphi_{n'\vec{k}}^{\sigma}|\varphi_{n\vec{k}}^{\sigma}\rangle\langle\varphi_{n\vec{k}}^{\sigma}|\varphi_{n'\vec{k}}^{\sigma}M_{J}\rangle}{\epsilon_{n\vec{k}}^{\sigma}-\epsilon_{n\vec{'k}}^{\sigma}}\label{eq: response matrix}\end{equation}
and

\begin{equation}
t_{I}^{\sigma}=2\sum_{\vec{k}}\sum_{n}^{\mathrm{occ.}}\sum_{n'}^{\mathrm{unocc.}}\langle\varphi_{n\vec{k}}^{\sigma}|V_{\mathrm{x,\mathrm{NL}}}^{\sigma}|\varphi_{n'\vec{k}}^{\sigma}\rangle\frac{\langle M_{I}\varphi_{n'\vec{k}}^{\sigma}|\varphi_{n\vec{k}}^{\sigma}\rangle}{\epsilon_{n\vec{k}}^{\sigma}-\epsilon_{n'\vec{k}}^{\sigma}}\,.\label{eq: represenatation rhs t}\end{equation}
Both core and valence states are taken into account in the sums over
the occupied states in Eqs.~\eqref{eq: response matrix} and \eqref{eq: represenatation rhs t}.

As said before, solving Eq.~\eqref{eq: algebraic OEP} for the exchange
potential involves the matrix inversion of $\chi_{s,IJ}^{\sigma}$.
The Hohenberg and Kohn theorem guarantees that the response function
is invertible except for variations of the potential given by an addition
of a constant. The latter restriction gives rise to a constant eigenfunction
of $\chi_{s,IJ}^{\sigma}$ with eigenvalue $0$, which we eliminate
from the outset by orthogonalizing all MPB functions to a constant
function such that variations in the potential by a constant are excluded.

Equation \eqref{eq: represenatation rhs t} contains the matrix elements
of the nonlocal exchange potential {[}Eq.~\eqref{eq: matrix elements of V_x_NL}]
between occupied (core and valence) and unoccupied states. In a recent
publication, we described an efficient scheme to calculate the valence-valence
and valence-conduction matrix elements within the FLAPW method.\cite{PBE0-NonLocalExactExchangePotential}
For the present implementation, this scheme has been extended to the
core-conduction matrix elements. Furthermore, spatial and time-reversal
symmetries are exploited to restrict the $\vec{k}$-point sums to
the irreducible wedge of the Brillouin zone in Eqs.~\eqref{eq: response matrix}
and \eqref{eq: represenatation rhs t}.\cite{PBE0-NonLocalExactExchangePotential,GW-MixedBasis}

\subsection{Numerical tests\label{sub: Tests}}

In this section, we present numerical tests of the spin-density response
function, the function $t_{I}^{\sigma}(\vec{r})$, and the resulting
exchange potential. According to the Eqs.~\eqref{eq: definition V_xc},
\eqref{eq: response function}, and \eqref{eq: rhs t} all three quantities
are functional derivatives of the form $\frac{\delta A(\vec{r})}{\delta B(\vec{r}')}$.
Thus, they describe the linear response of a quantity $A$ with respect
to changes of a quantity $B$. 

For the case of diamond, we calculate the changes $\Delta n(\vec{r})$
and $\Delta E_{\mathrm{x}}$ that result from an explicit perturbation
$V_{\mathrm{per}}(\vec{r})=\alpha\sum_{I}V_{\mathrm{per},I}M_{I}(\vec{r})$,
where $V_{\mathrm{per,}I}$ are random numbers, by exact diagonalization
of the perturbed Hamiltonian and check whether they correspond to
their linear counterpart $\Delta n^{\mathrm{lin}}(\vec{r})=\int\chi_{\mathrm{s}}(\vec{r},\vec{r}')V_{\mathrm{per}}(\vec{r}')d^{3}r'$,
$\Delta E_{\mathrm{x}}^{\mathrm{lin}}=\int t(\vec{r}')V_{\mathrm{per}}(\vec{r}')d^{3}r'$,
and $\Delta E_{\mathrm{x}}^{\mathrm{lin}}=\int V_{\mathrm{x}}(\vec{r}')\Delta n(\vec{r})d^{3}r$
up to linear order in $\alpha$.

Tables \ref{tab: Response function test}, \ref{tab: t test}, and
\ref{tab: vx test} show that, indeed, the differences $|\Delta n^{\mathrm{lin}}-\Delta n|$
and $|\Delta E_{\mathrm{x}}^{\mathrm{lin}}-\Delta E_{\mathrm{x}}|$
depend quadratically on the perturbation strength $\alpha$ which
confirms the validity of our implementation. %
\begin{table}

\caption{Numerical test for the response function $\chi_{\mathrm{s}}(\vec{r},\vec{r}')=\delta n(\vec{r})/\delta V_{\mathrm{s}}(\vec{r}')$
of diamond. The exact response of the density $\Delta n(\vec{r})$
is compared with its linear approximation $\Delta n^{\mathrm{lin}}(\vec{r})=\sum n_{I}^{\mathrm{lin}}M_{I}(\vec{r})$
with $n_{I}^{\mathrm{lin}}=\alpha\sum_{J}\chi_{\mathrm{s},IJ}V_{\mathrm{per,}J}$.
The L2 norm $[\int|\Delta n^{\mathrm{lin}}(\vec{r})-\Delta n(\vec{r})|^{2}d^{3}r]^{1/2}$clearly
shows a quadratic dependence on the perturbation strength $\alpha$.
\label{tab: Response function test}}

\begin{ruledtabular}

\begin{tabular}{clll}
$\alpha$&
0.01&
0.001&
0.0001\tabularnewline
\hline 
$|\Delta n^{\mathrm{lin}}$-$\Delta n|$&
$2.16710\times10^{-4}$&
$2.17018\times10^{-6}$&
$2.17035\times10^{-8}$\tabularnewline
\end{tabular}

\end{ruledtabular}

\end{table}
\begin{table}

\caption{Same as Table \ref{tab: Response function test} for $t(\vec{r})=\delta E_{\mathrm{x}}/\delta V_{\mathrm{s}}(\vec{r})$.
The difference of the exact response $\Delta E_{\mathrm{x}}$ of the
exchange energy and its linear approximation $\Delta E_{\mathrm{x}}^{\mathrm{lin}}=\alpha\sum_{I}t_{I}V_{\mathrm{per},I}$
depends quadratically on $\alpha$.\label{tab: t test}}

\begin{ruledtabular}

\begin{tabular}{clll}
$\alpha$&
0.01&
0.001&
0.0001\tabularnewline
\hline 
$\Delta E_{\mathrm{x}}$&
$1.69455$&
$1.74639\times10^{-1}$&
$1.75171\times10{}^{-2}$\tabularnewline
$\Delta E_{\mathrm{x}}^{\mathrm{lin}}$&
$1.75230$&
$1.75230\times10^{-1}$&
$1.75230\times10^{-2}$\tabularnewline
$|\Delta E_{\mathrm{x}}^{\mathrm{lin}}$-$\Delta E_{\mathrm{x}}|$&
$5.77479\times10^{-2}$&
$5.90882\times10^{-4}$&
$5.92211\times10^{-6}$\tabularnewline
\end{tabular}

\end{ruledtabular}

\end{table}
\begin{table}

\caption{Same as Table \ref{tab: t test}, for $V_{\mathrm{x}}(\vec{r})=\delta E_{\mathrm{x}}/\delta n(\vec{r})$
with $\Delta E_{\mathrm{x}}^{\mathrm{lin}}=\sum_{I}V_{\mathrm{x},I}\int M_{I}^{*}(\vec{r})\Delta n(\vec{r})d^{3}r$.\label{tab: vx test}}

\begin{ruledtabular}

\begin{tabular}{clll}
$\alpha$&
0.01&
0.001&
0.0001\tabularnewline
\hline 
$\Delta E_{\mathrm{x}}$&
$1.69455$&
$1.74639\times10^{-1}$&
$1.75171\times10^{-2}$\tabularnewline
$\Delta E_{\mathrm{x}}^{\mathrm{lin}}$&
$1.67924$&
$1.74479\times10^{-1}$&
$1.75155\times10^{-2}$\tabularnewline
$|\Delta E_{\mathrm{x}}^{\mathrm{lin}}$-$\Delta E_{\mathrm{x}}|$&
$1.53093\times10^{-2}$&
$1.59919\times10^{-4}$&
$1.60625\times10^{-6}$\tabularnewline
\end{tabular}

\end{ruledtabular}

\end{table}

\section{Results \& Discussion\label{sec: Results-&-Discussion}}

In this section, we present results for a variety of semiconductors
and insulators obtained with our implementation of the EXX-OEP approach
within the FLAPW method. In particular, we demonstrate for the case
of diamond that a smooth and physical EXX potential requires a balance
of the basis sets: the LAPW basis for the wave functions must be converged
with respect to a given MPB until the EXX potential does not change
anymore. This is somewhat counterintuitive and in contrast to our
implementation of the hybrid functionals where, conversely, the MPB
must be converged for a given LAPW basis. A similar behavior has been
found in implementations employing plane-wave and Gaussian basis sets.\cite{Balance-Gaussian,Hesselmann}
We will analyze and explain this point later in this section.

\begin{figure}
\includegraphics[clip,scale=0.35,angle=-90]{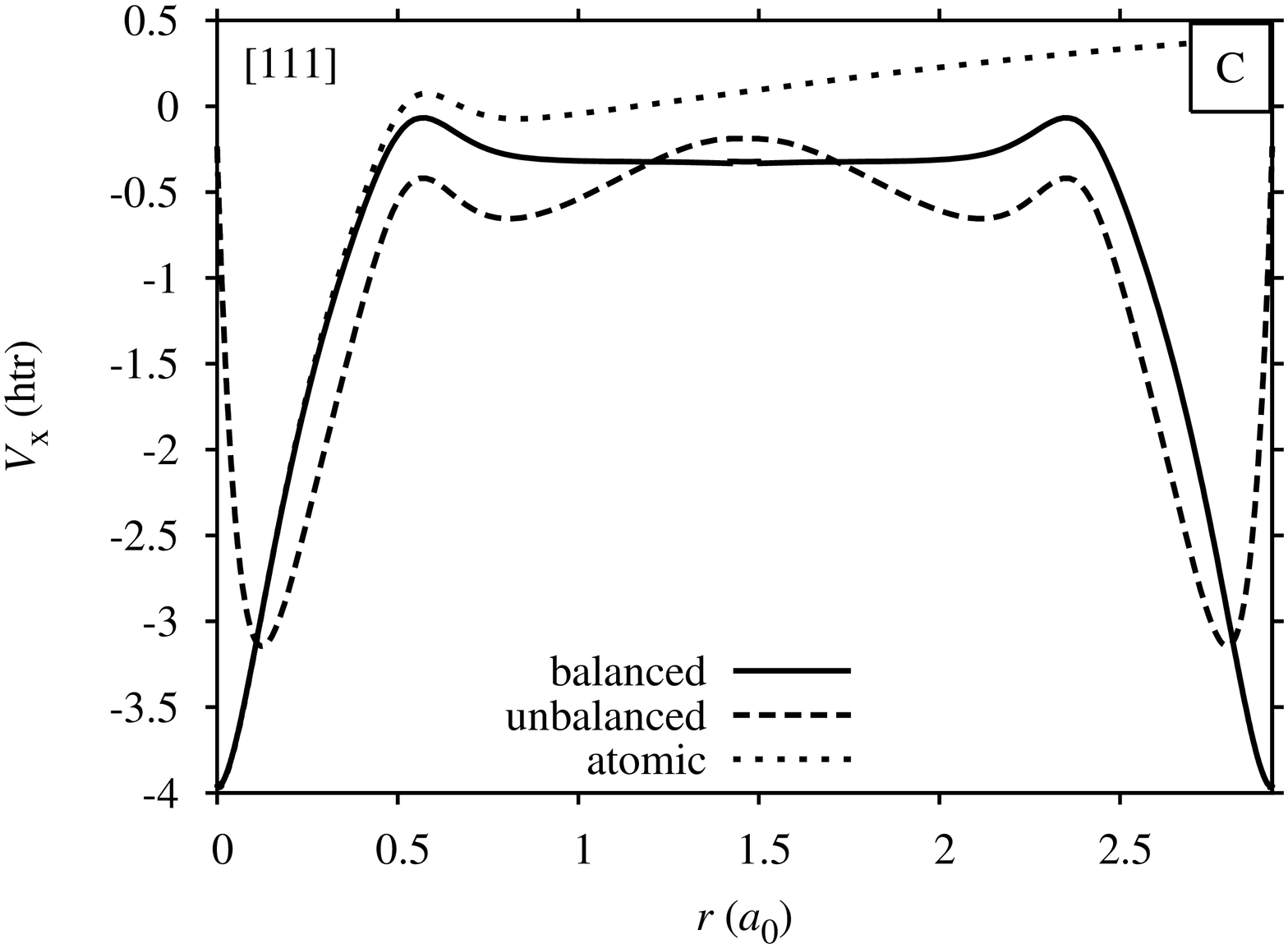}

\begin{raggedright}(a)\par\end{raggedright}

\includegraphics[clip,scale=0.35,angle=-90]{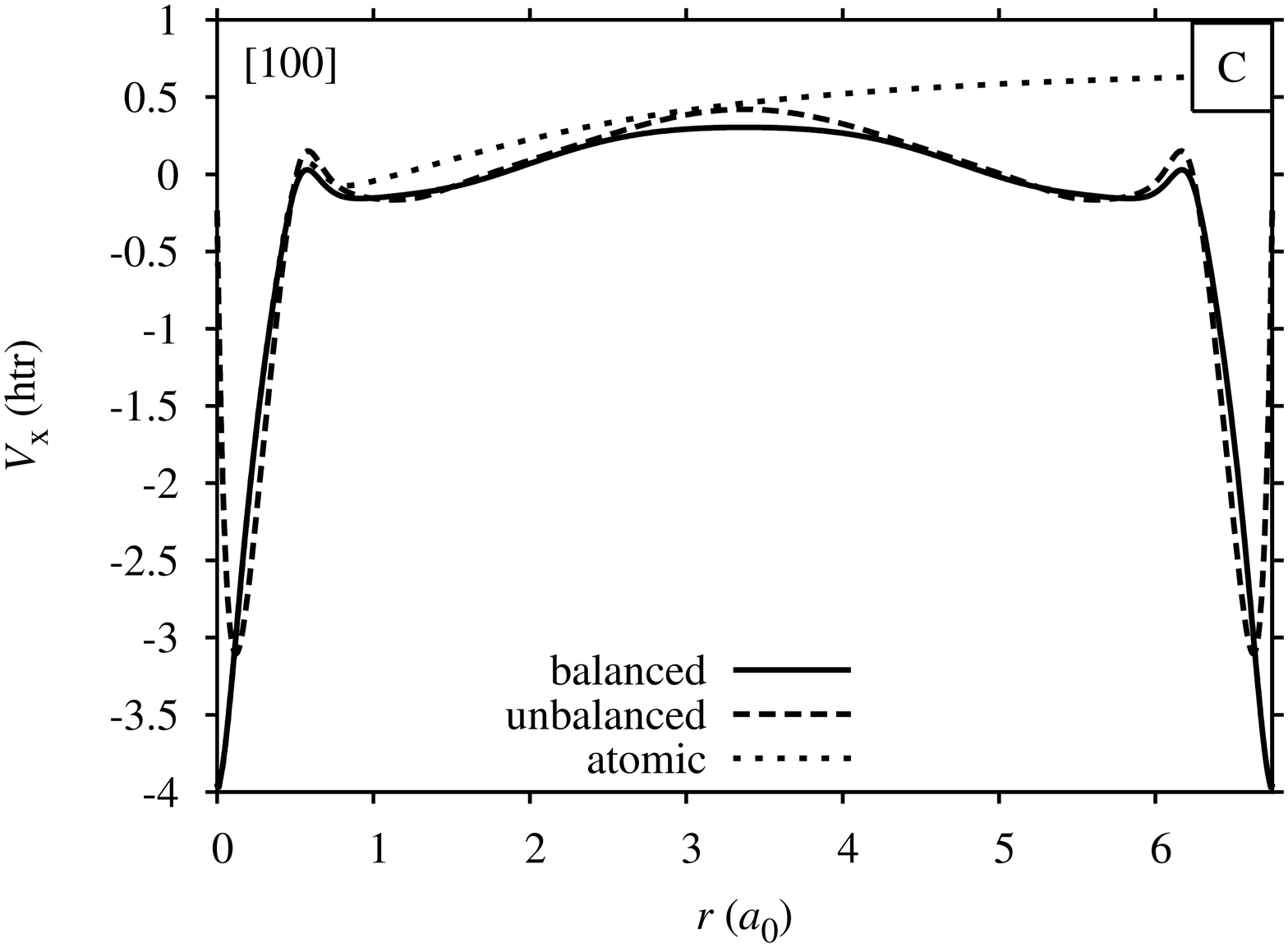}

\begin{raggedright}(b)\par\end{raggedright}

\caption{Local EXX potential of diamond on lines starting and ending at atomic
nuclei along (a) the {[}111] and (b) the {[}100] directions. Dashed
and solid lines correspond to cases where the MPB and LAPW basis sets
are \textit{unbalanced} and \textit{balanced}, respectively. For comparison,
we also show the atomic EXX potential of carbon as dotted lines, shifted
to align with the crystal EXX potential at the atomic nucleus at $r=0$.
\label{fig: C EXX potential (L=3D4)}}
\end{figure}

\begin{figure}
\includegraphics[clip,scale=0.2]{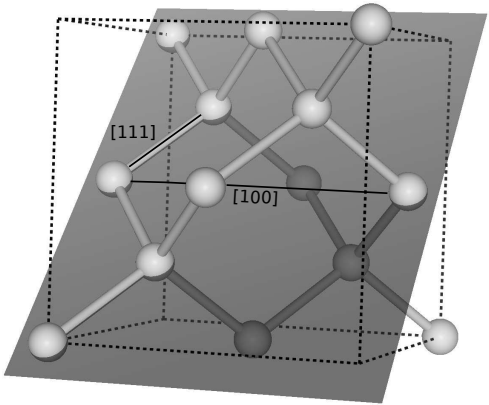}

\caption{Cubic unit cell of diamond with the ($01\overline{1}$) plane. The
lines connecting two carbon atoms along the $[111]$ and $[100]$
directions are indicated as solid lines. \label{fig: C unit-cell}}
\end{figure}

Figure~\ref{fig: C EXX potential (L=3D4)} shows the local EXX potential
on lines connecting two neighboring carbon atoms along the $[111]$
{[}Fig.~\ref{fig: C EXX potential (L=3D4)}(a)] and the $[100]$
{[}Fig.~\ref{fig: C EXX potential (L=3D4)}(b)] directions, see Fig.~\ref{fig: C unit-cell}.
A $4$$\times$$4$$\times$$4$ $\vec{k}$-point sampling is employed,
and the MPB parameters are $G'_{\mathrm{max}}=3.4\, a_{0}^{-1}$ ($a_{0}$
is the Bohr radius) and $L_{\mathrm{max}}=4$, giving rise to five
$s$\hbox{-}, four $p$\hbox{-}, four $d$\hbox{-}, and three $f$\hbox{-},
and two $g$-type radial functions per atom. These cutoff values are
well below those of the LAPW basis, $l_{\mathrm{max}}=6$ and $G_{\mathrm{max}}=4.2\, a_{0}^{-1}$,
which reflects the relative smoothness of the potential compared with
the shape of the wave functions. However, if we only use the conventional
basis of augmented plane waves, defined in Eq.~\eqref{eq: APWs},
the potential (dashed lines) shows an overpronounced intershell hump
and tends to an unphysical positive value close to the atomic nucleus
($r=0$). This is a case where the basis sets are \textit{unbalanced}.
In particular, the LAPW basis lacks flexibility in the MT spheres
as becomes obvious when we add local orbitals, which are nonzero only
in the MT spheres. We find that six local orbitals for each $lm$
channel with $l=0,\dots,5$ and $|m|\le l$, placed at higher energies
according to the prescription described in Sec.~\ref{sec: FLAPW-Method},
are needed to converge the local EXX potential. This is reasonable
since, with the cutoff $L_{\mathrm{max}}=4$, the occupied $2s$ and
$2p$ states of diamond couple maximally to the $l=5$ contribution
of the unoccupied states. With so many local orbitals the number of
basis functions is increased roughly by a factor of five: there are
about $100$ augmented plane waves (the exact number depends on the
$\vec{k}$ point) and $432$ additional local orbitals. All resulting
KS bands, about $530$, are taken into account in the sums of Eqs.~\eqref{eq: response explicit}
and \eqref{eq: rhs t explicit}. The resulting potential is shown
as solid lines in Fig.~\ref{fig: C EXX potential (L=3D4)} and looks
smooth and physical. It is remarkable that, even for diamond, it takes
so much effort to converge the EXX potential since, in conventional
LDA or GGA calculations, diamond is treated readily with a very modest
LAPW basis without any local orbitals.%

Before analyzing this point in more detail, we want to identify the
MPB functions that contribute most to the MT part of the potential.
Not surprisingly, the function that corresponds to the atomic EXX
potential gets the largest weight. In fact, close to the atomic nuclei
this function (dotted lines in Fig.~\ref{fig: C EXX potential (L=3D4)})
and its bulk counterpart are indistinguishable. They deviate more
and more towards the MT sphere boundary ($S\mathrm{=1.42\,}a_{0}$),
where the atomic EXX potential already enters the typical $1/r$ behavior,
while the crystal EXX potential is periodic. The second largest contribution
comes from the constant MT function, which helps to align the MT potential
to the interstitial one. We note that there is no ambiguity with respect
to adding a constant to the potential over the whole space, since
the constant function has been eliminated explicitly from the MPB
(see Sec.~\ref{sub: Single-particle-response}) giving rise to the
condition $\int V_{\mathrm{x}}(\vec{r})d^{3}r=0$. 

\begin{figure}
\includegraphics[scale=0.35,angle=-90]{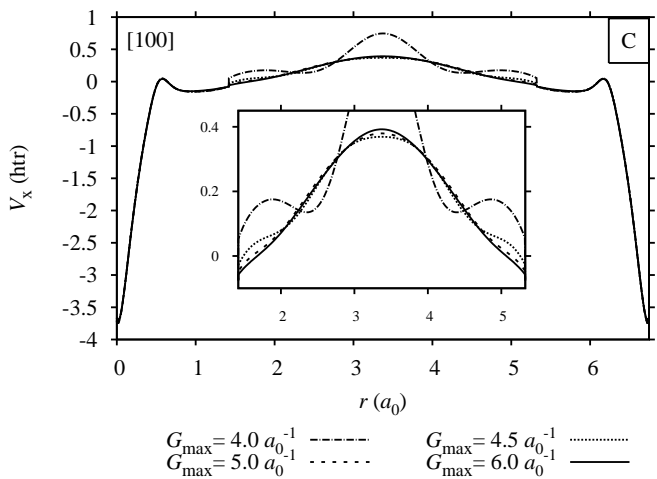}

\caption{Convergence of the interstitial EXX potential on a line connecting
two carbon atoms along the {[}100] direction for four different reciprocal
cutoff radii of the LAPW basis. The corresponding MPB cutoff is fixed
to $5.8\, a_{0}^{-1}$. \label{fig: IR balance}}
\end{figure}

So far, we have only discussed the MT potential, whose proper convergence
requires additional local orbitals in the spheres. We find an analogous
behavior for the interstitial potential. As is seen in Fig.~\ref{fig: IR balance},
the cutoff radius of the reciprocal lattice vectors included in the
LAPW basis set must be converged with respect to that of the MPB.
To show this effect clearly, the latter was chosen much larger than
necessary, $G'_{\mathrm{max}}=5.8\, a_{0}^{-1}$. Similarly to the
MT potential, the interstitial potential exhibits spurious oscillations
in the underconverged cases. Only if $G_{\mathrm{max}}$ is large
enough, $G_{\mathrm{max}}\gtrsim5.0\, a_{0}^{-1}$, the oscillations
are suppressed and a smooth potential is obtained. Furthermore, Fig.~\ref{fig: IR balance}
shows that, in the underconverged cases, the potential is not continuous
at the MT sphere boundary because the oscillatory potentials possess
large-$\vec{G}$ Fourier coefficients and require spherical harmonics
beyond $L_{\mathrm{max}}=4$ in the spheres for a proper matching.
Fortunately, the converged EXX potential is a smooth function and
already moderate reciprocal cutoff radii are sufficient, typically
75\% of the usual LAPW cutoff. For diamond, for example, the combination
of $G'_{\mathrm{max}}=3.4\, a_{0}^{-1}$ and $G_{\mathrm{max}}=4.2\, a_{0}^{-1}$
leads to stable results.%

\begin{figure}[!h]
(a)\includegraphics[clip,scale=0.87]{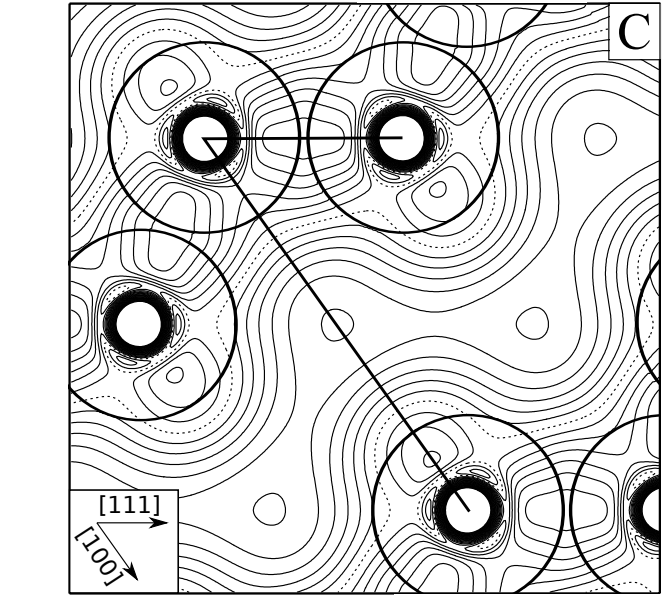}

(b)\includegraphics[clip,scale=0.87]{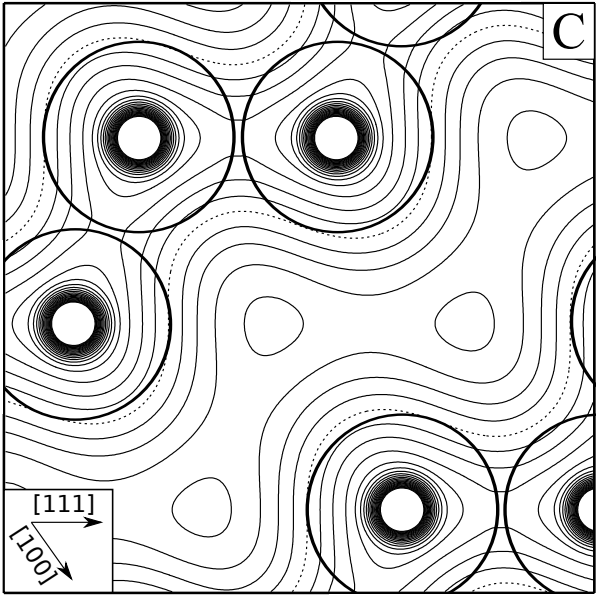}

\caption{Two-dimensional plot of (a) the local EXX and (b) the LDA exchange
potential on the $(01\overline{1})$ plane of diamond. Both potentials
are shifted such that $\int V_{\mathrm{x}}(\vec{r})d^{3}r=0$. The
contour lines start at $V_{\mathrm{x}}=-1.00\,\mathrm{htr}$ and have
an interval of $0.05\,\mathrm{htr}$. The contour line with the maximal
value is found (a) at $V_{\mathrm{x}}=0.30\,\mathrm{htr}$ and (b)
at $V_{\mathrm{x}}=0.20\,\mathrm{htr}$. The dotted line corresponds
to $V_{\mathrm{x}}=0$. The MT sphere boundaries and the lines corresponding
to Fig.~\ref{fig: C EXX potential (L=3D4)} are indicated.\label{fig: contour plot}}
\end{figure}

\begin{figure}[!h]
(a)\includegraphics[clip,scale=0.87]{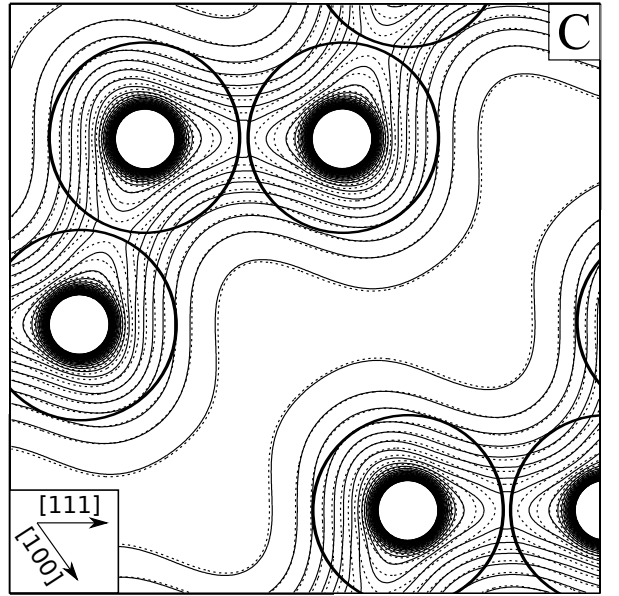}

(b)\includegraphics[bb=10bp 35bp 297bp 315bp,clip,scale=0.87]{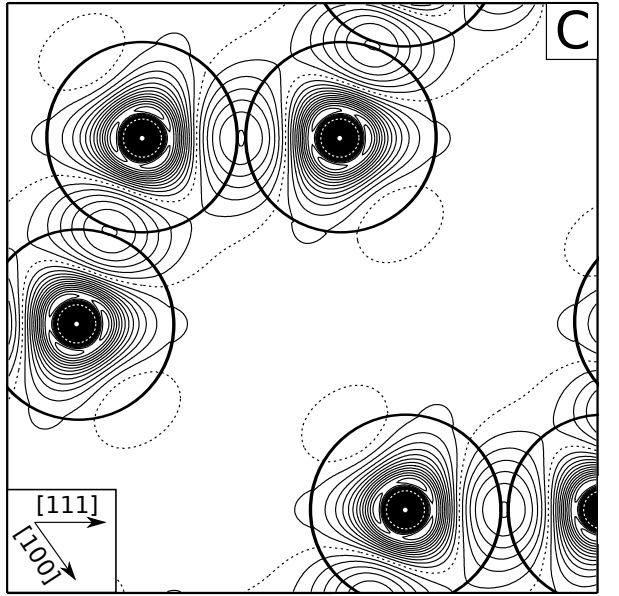}

\caption{(a) Total electron densities obtained from the EXX (solid lines)
and LDA potential (dotted lines) on the diamond $(01\overline{1})$
plane. Contour lines for the values $0.03\, a_{0}^{-3},\,0.06\, a_{0}^{-3},\dots,\,0.99\, a_{0}^{-3}$
are shown. (b) Plot of the density difference with contour lines between
$-0.039\, a_{0}^{-3}$ and $0.390\, a_{0}^{-3}$ at intervals of $0.003\, a_{0}^{-3}$.
The dotted line corresponds to $\Delta n(\vec{r})=0$. \label{fig: density-contour plot}}
\end{figure}

An overall view of the MT and the interstitial exchange potential
on the diamond ($01\overline{1}$) plane is shown in Fig.~\ref{fig: contour plot}(a)
as a contour plot. The plane is displayed in Fig.~\ref{fig: C unit-cell}.
It contains the connecting lines along {[}111] and {[}100] that correspond
to Fig.~\ref{fig: C EXX potential (L=3D4)}. We see that in the regions
close to the atomic nuclei the potential is predominantly spherical.
However, towards the MT sphere boundaries the potential becomes strongly
anisotropic and matches continuously to the warped interstitial potential,
which is far from constant also. In fact, the nonsphericity of the
EXX potential is considerably more pronounced than in the LDA potential,
Fig.~\ref{fig: contour plot}(b). The latter is similar in shape
to the electron density distribution {[}cf.~Fig.~\ref{fig: density-contour plot}(a)],
of which it is a direct function $V_{\mathrm{x}}^{\mathrm{LDA}}(\vec{r})\propto n(\vec{r})^{1/3}$.
The EXX potential, in contrast, incorporates the full nonlocality
of the EXX functional, which makes it much more corrugated than the
LDA one, in particular in the MT spheres, where the KS orbitals are
highly oscillatory. All this stresses the importance of a full-potential
treatment within the EXX-OEP approach.

The LDA potential corresponds in each point $\vec{r}$ to the exchange
potential of the homogeneous electron gas with an electron density
that equals the local electron density $n(\vec{r})$ of the real system.
Thus, by construction it is exact for the homogeneous electron gas
but misses the effects of density variations. The EXX potential, in
contrast, takes all density variations exactly into account. Thus,
the differences between Figs.~\ref{fig: contour plot}(a) and (b)
must be attributed to the influence of the density inhomogeneities
on the exchange potential. This influence is particularly large in
regions where the density varies a lot, that is, close to the atomic
nuclei, while in the interstitial region the two potentials are more
similar.

The differences in the exchange potentials naturally affect the electron
density distribution. The EXX electron density, Fig.~\ref{fig: density-contour plot}(a),
clearly shows a pronounced contraction of the electron distribution
compared with the LDA one. This is a direct consequence of the self-interaction
error, which is eliminated in the EXX approach, while it gives rise
to an unphysical delocalization in the case of the LDA potential.
This becomes clearer in Fig.~\ref{fig: density-contour plot}(b)
where we plot the difference between the EXX and the LDA densities.
The exactly compensated self-interaction allows the charge to accumulate
in the atomic cores, but also in the covalent bonds between the atoms. 


To understand the requirement of the basis-set balance in more detail,
we go back to the OEP Eq.~\eqref{eq: algebraic OEP}, whose solution
involves the inversion of the response function, Eq.~\eqref{eq: response matrix}.
The response function describes the linear response of the electron
density with respect to changes of the effective potential. For the
latter we employ the MPB, while the former is given by the orbital
densities of the occupied states, Eq.~\eqref{eq: density}, and,
hence, ultimately by the LAPW basis set. Thus, the LAPW basis must
provide enough flexibility for the density to enable it to respond
adequately to the changes of the effective potential.

\begin{figure}
\includegraphics[scale=0.34,angle=-90]{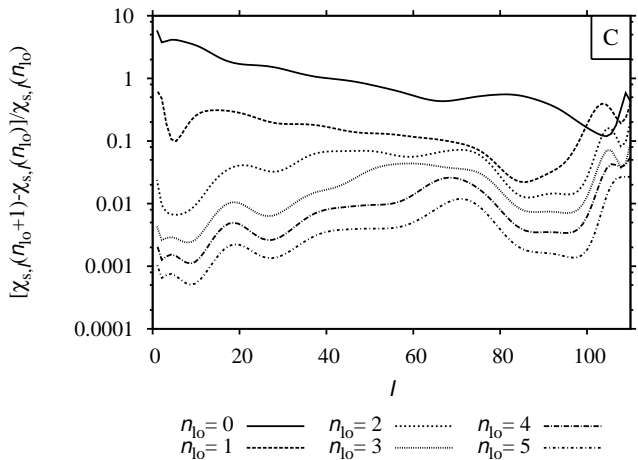}

\caption{Convergence of the eigenvalues $\chi_{\mathrm{s},I}$ of the response
function for diamond. The $\chi_{\mathrm{s},I}$ are all negative
and ordered according to increasing moduli. The relative change $[\chi_{\mathrm{s},I}(n_{\mathrm{lo}}+1)-\chi_{\mathrm{s},I}(n_{\mathrm{lo}})]/\chi_{\mathrm{s},I}(n_{\mathrm{lo}})$
is plotted, where $n_{\mathrm{lo}}$ is the number of local orbitals
per $lm$ channel ($l=0,\dots,5$, $|m|\le l$). The case $n_{\mathrm{lo}}=0$
corresponds to the conventional LAPW basis without local orbitals.
A Bezier algorithm was employed to smooth the curves.\label{fig: Response convergence}}
\end{figure}

This explains the observed behavior and becomes evident in the convergence
of the response function with respect to the LAPW basis. In Fig.~\ref{fig: Response convergence}
we show the changes of the eigenvalues of $\chi_{\mathrm{s}}$, ordered
according to increasing moduli, when we add more and more local orbitals,
$2l+1$ additional local orbitals per $l$ quantum number ($l=0,\dots,5$)
in each step. We again use $L_{\mathrm{max}}=4$ and $G'_{\mathrm{max}}=3.4\, a_{0}^{-1}$
as MPB cutoff values. In the case of the maximal number of local orbitals
per $lm$ channel, $n_{\mathrm{lo}}=6$, the basis is increased by
$432$ functions relative to the conventional LAPW basis. Clearly,
the eigenvalues can be systematically converged. The relative changes
are in the order of 0.1\% to 1.0\% between $n_{\mathrm{lo}}=5$ and
$n_{\mathrm{lo}}=6$. Especially the small eigenvalues converge well,
which are particularly important in the inversion of the response
function. We note that a straightforward elimination of the small
eigenvalues by singular value decomposition is not advisable and leads
to an ill-defined response function. %

Up to now, we have discussed the importance of the quality of the
LAPW basis for the shape of the local EXX potential. Only for well-balanced
basis sets, the potential is smooth and physical (cf.~Fig.~\ref{fig: C EXX potential (L=3D4)}).
We now address the question to what extent this effect influences
the KS one-particle energies that result from the self-consistent
solution of Eq.~\eqref{eq: KS Hamiltonian} with the local EXX potential
obtained from the OEP Eq.~\eqref{eq: x OEP equation}. Table \ref{tab: convergence of transition energies}
gives the transition energies, that is, the KS eigenvalue differences,
from the valence-band maximum at the $\Gamma$ point to the $\mathrm{L}$
and $\mathrm{X}$ point of the lowest conduction band of diamond for
the basis sets with $n_{\mathrm{lo}}=0,\dots,6$ local orbitals per
$lm$ channel. Obviously, at least three local orbitals are necessary
to converge these transition energies to within $0.01\,\mathrm{eV}$.
Between the unbalanced and balanced basis sets ($n_{\mathrm{lo}}=0$
and $n_{\mathrm{lo}}=6$, respectively) the values change by about
0.2 eV.%
\begin{table}
\begin{ruledtabular}

\begin{tabular}{cccc}
$n_{\mathrm{lo}}$&
$\Gamma\rightarrow\Gamma$&
$\Gamma\rightarrow\mathrm{L}$&
$\Gamma\rightarrow\mathrm{X}$\tabularnewline
\hline
0&
6.351&
9.243&
5.307\tabularnewline
1&
6.196&
9.086&
5.125\tabularnewline
2&
6.186&
9.069&
5.144\tabularnewline
3&
6.180&
9.063&
5.138\tabularnewline
4&
6.178&
9.059&
5.139\tabularnewline
5&
6.177&
9.057&
5.136\tabularnewline
6&
6.176&
9.055&
5.136\tabularnewline
\end{tabular}

\end{ruledtabular}

\caption{KS transition energies (in eV) for diamond obtained from the self-consistent
solution of the KS equation with the local EXX potential for LAPW
basis sets including zero to six local orbitals per $lm$ channel
($l=0,\dots,5$, $|m|\le l$). As the LAPW basis is made more flexible,
the transition energies converge. \label{tab: convergence of transition energies}}
\end{table}

\begin{figure}
\includegraphics[scale=0.32,angle=-90]{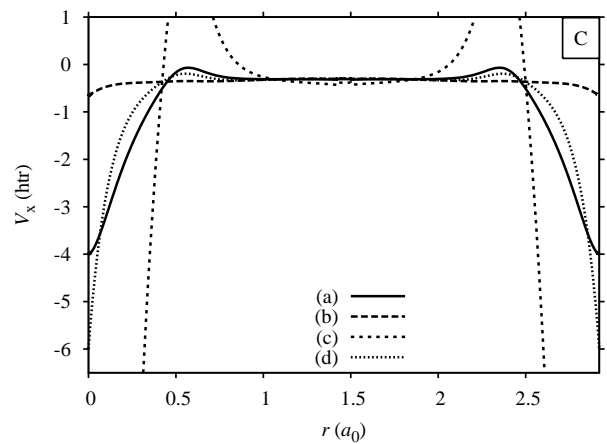}

\caption{Same as Fig.~\ref{fig: C EXX potential (L=3D4)} for the cases (a)
$1s$ state fully considered, (b) only considered in $\chi_{\mathrm{s},IJ}$,
(c) only considered in $t_{I}$, and (d) neglected in both (see text).\label{fig: balance left/right}}
\end{figure}

Another balance condition we find for the $1s$ core state that goes
into both the left- and the right-hand sides of Eq.~\eqref{eq: algebraic OEP}.
In particular, we now distinguish between four cases: (a) core state
considered in $\chi_{\mathrm{s},IJ}$ and $t_{I}$, (b) core state
only considered in $\chi_{\mathrm{s},IJ}$, (c) core state only considered
in $t_{I}$, and (d) core state neglected in both. Case (a) corresponds
to the full calculations presented so far. Figure \ref{fig: balance left/right}
shows that the resulting potentials look very different for the different
cases. Surprisingly, potential (d) is closest to the full potential,
while the potentials (b) and (c) are much too shallow and too strongly
varying, respectively. Obviously, the inclusion of the core state
only on one side of the OEP equation gives rise to an equation that
is \textit{out of balance} and that yields an unphysical $V_{\mathrm{x}}(\vec{r})$.
This also influences the resulting KS transition energies. In the
balanced case (d), the results $6.19\,\mathrm{eV}$, $9.08\,\mathrm{eV}$,
and $5.28\,\mathrm{eV}$ for $\Gamma\rightarrow\Gamma$, $\Gamma\rightarrow\mathrm{L}$,
and $\Gamma\rightarrow\mathrm{X}$, respectively, are surprisingly
close to the full calculation (a) (cf.~Table \ref{tab: convergence of transition energies}),
while the energies for the unbalanced case (b) deviate more strongly,
especially for the $\Gamma\rightarrow\mathrm{X}$ transition, $6.22\,\mathrm{eV}$,
$9.16\,\mathrm{eV}$, and $6.03\,\mathrm{eV}$. The energetic position
of the $1s$ state with respect to the Fermi energy, however, is only
realistic for the full calculation (a), $\epsilon(1s)=-265.7\,\mathrm{eV}$
(cf.~$-262.9\,\mathrm{eV}$ for LDA), whereas (b) and (d) give binding
energies, that are smaller by $31.6\,\mathrm{eV}$ and $10.5\,\mathrm{eV}$,
respectively. Calculation (c) is unstable and does not converge. These
results indicate that the PP approximation is, indeed, suitable for
the EXX-OEP approach, a conjecture that will be confirmed by our reference
calculations later-on.%

As outlined in Sec.~\ref{sec: Theory}, the construction of a local
EXX potential within the KS formalism of DFT is equivalent to the
OEP approach of Sharp and Horton,\cite{OEP/OPM} where the Hartree-Fock
total energy is minimized under the constraint that the wave functions
feel a local multiplicative potential. This constraint reduces the
Hilbert space for the wave functions and, thus, increases the total
energy due to the variational principle. In fact, we find that the
total energy for diamond obtained with the nonlocal Hartree-Fock potential,
Eq.~\eqref{eq: nonlocal HF kernel}, and an $8$$\times$$8$$\times$$8$
$\vec{k}$-point set is $0.24\,\mathrm{eV}$ per unit cell lower than
that of the EXX-OEP approach. 

As reference, we now report fully converged KS transition energies
for a variety of semiconductors and insulators in Table \ref{tab: EXX+EXX-VWN results}.
All calculations are performed with an $8$$\times$$8$$\times$$8$
Brillouin zone sampling and at the experimental lattice constants
(C:~$6.746\, a_{0}$, Si:~$10.26\, a_{0}$, SiC:~$8.24\, a_{0}$,
Ge:~$10.67\, a_{0}$, GaAs:~$10.68\, a_{0}$, Ne:~$8.44\, a_{0}$,
Ar:~$9.93\, a_{0}$). Apart from the EXX-only calculations, we also
show transition energies obtained with the EXX+VWNc functional in
which the LDA correlation functional from Ref.~\onlinecite{LDA-VWN}
was added to the EXX functional. Table \ref{tab: EXX+EXX-VWN results}
shows that both the EXX and EXX+VWNc functionals yield KS transition
energies much closer to experiment (last column) than the LDA functional
(first column). While for semiconductors the resulting energies are
even quantitatively in very good agreement with experiment, there
are larger discrepancies for the insulators diamond and, in particular,
crystalline Ne and Ar. There is only little difference between the
EXX and EXX+VWNc values. The inclusion of the LDA correlation functional
does not lead to a definite improvement. The direct band gap $\Gamma\rightarrow\Gamma$
for diamond of $6.21\,\mathrm{eV}$ agrees very well with the value
$6.18\,\mathrm{eV}$ recently reported by Engel,\cite{Engel} who
used a plane-wave PP approach with cutoff values pushed to the AE
limit. It is even identical to the value obtained with a standard
valence-only plane-wave PP approach.\cite{Hesselmann-privat} In contrast
to that, Sharma \textit{et al.} calculated a much larger value of
$6.67\,\mathrm{eV}$ with their FLAPW-EXX implementation. For neon,
there is a somewhat larger discrepancy with the calculation by Magyar
\textit{et al.},\cite{EXX-Ar/Ne} though. In conclusion, with our
EXX-OEP implementation within the AE FLAPW method, we obtain results
in very good agreement with previous plane-wave PP calculations, provided
that the basis sets for the wave functions and the potential are properly
balanced. This shows that the PP approximation is adequate for the
EXX-OEP approach at least for the systems examined here, which is
at variance with the findings of Ref.~\onlinecite{Sharma}.%
\begin{table*}
\begin{ruledtabular}

\begin{tabular}{cccccccc}
&
&
\multicolumn{3}{c}{This work}&
\multicolumn{2}{c}{Plane-wave PP}&
\tabularnewline
\cline{3-5} \cline{6-7} 
&
&
LDA&
EXX&
EXX+VWNc&
EXX&
EXX+VWNc&
Expt.\tabularnewline
\hline
C&
$\Gamma\rightarrow\Gamma$&
$5.56$&
$6.21$&
$6.26$&
$6.19^{a},6.21^{b}$&
$6.28^{c}$&
$7.3^{e}$\tabularnewline
&
$\Gamma\rightarrow\mathrm{L}$&
$8.43$&
$9.09$&
$9.16$&
$9.15^{b}$&
$9.18^{c}$&
\tabularnewline
&
$\Gamma\rightarrow\mathrm{X}$&
$4.71$&
$5.20$&
$5.33$&
$5.34^{b}$&
$5.43^{c}$&
\tabularnewline
Si&
$\Gamma\rightarrow\Gamma$&
$2.53$&
$3.13$&
$3.21$&
$3.12^{b}$&
$3.26^{c}$&
$3.4^{e}$\tabularnewline
&
$\Gamma\rightarrow\mathrm{L}$&
$1.42$&
$2.21$&
$2.28$&
$2.21^{b}$&
$2.35^{c}$&
$2.4^{e}$\tabularnewline
&
$\Gamma\rightarrow\mathrm{X}$&
$0.61$&
$1.30$&
$1.44$&
$1.25^{b}$&
$1.50^{c}$&
\tabularnewline
SiC&
$\Gamma\rightarrow\Gamma$&
$6.27$&
$7.18$&
$7.24$&
&
$7.37^{c}$&
\tabularnewline
&
$\Gamma\rightarrow\mathrm{L}$&
$5.38$&
$6.14$&
$6.21$&
&
$6.30^{c}$&
\tabularnewline
&
$\Gamma\rightarrow\mathrm{X}$&
$1.32$&
$2.29$&
$2.44$&
&
$2.52^{c}$&
$2.42^{e}$\tabularnewline
Ge&
$\Gamma\rightarrow\Gamma$&
$-0.14$&
$1.24$&
$1.21$&
&
$1.28^{c}$&
$1.0^{e}$\tabularnewline
&
$\Gamma\rightarrow\mathrm{L}$&
$0.06$&
$0.89$&
$0.94$&
&
$1.01^{c}$&
$0.7^{e}$\tabularnewline
&
$\Gamma\rightarrow\mathrm{X}$&
$0.66$&
$1.15$&
$1.28$&
&
$1.34^{c}$&
$1.3^{e}$\tabularnewline
GaAs&
$\Gamma\rightarrow\Gamma$&
$0.29$&
$1.72$&
$1.74$&
&
$1.82^{c}$&
$1.63^{e}$\tabularnewline
&
$\Gamma\rightarrow\mathrm{L}$&
$0.85$&
$1.79$&
$1.86$&
&
$1.93^{c}$&
\tabularnewline
&
$\Gamma\rightarrow\mathrm{X}$&
$1.35$&
$1.95$&
$2.12$&
&
$2.15^{c}$&
$2.18^{e}$\tabularnewline
Ne&
$\Gamma\rightarrow\Gamma$&
$11.43$&
$14.79$&
$15.46$&
$14.15^{d}$&
$14.76^{d}$&
$21.51^{f}$\tabularnewline
&
$\Gamma\rightarrow\mathrm{L}$&
$16.97$&
$20.49$&
$21.16$&
&
&
\tabularnewline
&
$\Gamma\rightarrow\mathrm{X}$&
$18.27$&
$21.85$&
$22.56$&
&
&
\tabularnewline
Ar&
$\Gamma\rightarrow\Gamma$&
$8.19$&
$9.65$&
$10.09$&
$9.61^{d}$&
$9.95^{d}$&
$14.15^{f}$\tabularnewline
&
$\Gamma\rightarrow\mathrm{L}$&
$11.06$&
$12.22$&
$12.60$&
&
&
\tabularnewline
&
$\Gamma\rightarrow\mathrm{X}$&
$10.86$&
$12.08$&
$12.49$&
&
&
\tabularnewline
\end{tabular}

\end{ruledtabular}

\begin{flushleft}\begin{tabular}{llllll}
$^{a}$Reference &
\onlinecite{Engel}&
$^{b}$Reference &
\onlinecite{Hesselmann-privat}&
$^{c}$Reference&
\onlinecite{EXX-PP-Staedele-2}\tabularnewline
$^{d}$Reference &
\onlinecite{EXX-Ar/Ne}&
$^{e}$Reference &
\onlinecite{Landolt-Boernstein}&
$^{f}$Reference&
\onlinecite{Ne-Ar}\tabularnewline
\end{tabular}\end{flushleft}

\caption{KS transition energies (in eV) obtained with the local EXX and EXX+VWNc
potentials and an $8$$\times$$8$$\times$$8$ $\vec{k}$-point
sampling. For comparison, plane-wave PP results and experimental values
from the literature are given.\label{tab: EXX+EXX-VWN results}}
\end{table*}

\section{Conclusions\label{sec: Conclusions}}

We have developed an all-electron full-potential implementation of
the EXX-OEP approach to DFT within the FLAPW method. We analysed the
conditions and requirements on the basis sets and numerical cutoff
parameters to obtain reliable and numerically stable results. Based
on this knowledge we presented as proof of principle results on KS
transition energies for some typical semiconductors, insulators and
noble-gas solids that are in very good agreement with pseudopotential
results.

The OEP equation is formulated utilizing the mixed product basis (MPB),\cite{PBE0-NonLocalExactExchangePotential,GW-MixedBasis,CoulombMatrix-MixedBasis}
which has been adjusted for the present purpose: it is augmented with
the atomic EXX potential, the constant function is eliminated, and
the basis functions are made continuous all over the space. In this
basis, the OEP equation becomes an algebraic equation, which is solved
for the local EXX potential with standard numerical tools.

For the case of diamond, we have demonstrated that the local EXX potentials
are spatially strongly corrugated, which makes a full-potential treatment
even more important than in conventional LDA or GGA calculations.
Furthermore, the two basis sets, LAPW and MPB are not independent.
They must be properly balanced to obtain a smooth and physical EXX
potential over the whole space. In the \textit{unbalanced} case, the
potential shows spurious oscillations, which we have traced back to
an insufficiently converged response function, a function that gives
the response of the electron density with respect to changes of the
effective potential. If the LAPW basis, which parametrizes the electron
density, is not flexible enough, the electron density cannot follow
the changes of the effective potential that are described by the MPB
leading to a corrupted response function. As a result, the LAPW basis
must be converged with respect to a given MPB. Already in the simple
case of diamond, we must add six local orbitals at different energies
for each $lm$ channel from $l=0$ to $l=5$ in order to obtain a
smooth potential in the spheres. This shows that the LAPW basis must
be converged to an accuracy that is far beyond that of conventional
LDA or GGA calculations. Similarly, also the LAPW reciprocal cutoff
radii must be chosen large enough.

Not surprisingly, the shape of the EXX potential -- oscillatory or
smooth -- has an impact on the resulting KS transition energies. We
find that with properly balanced basis sets, the transition energies
for a variety of semiconductors and insulators obtained with the EXX
and the EXX+VWNc functionals are in very good agreement with plane-wave
pseudopotential results from the literature (crystalline neon is an
exception). This confirms that the pseudopotential approximation works
reliably within the EXX-OEP approach. Our finding is in contradiction
to a previously published implementation (Ref.~\onlinecite{Sharma})
based on the FLAPW method, where large discrepancies with pseudopotential
results were reported.

Currently, reliable all-electron full-potential EXX-OEP calculations
are computationally very demanding, because of the need for large
orbital basis sets, which we attribute partly to the fact that the
LAPW basis functions depend explicitly on the effective potential.
To refine our full-potential implementation of the EXX-OEP approach,
we suggest as an task for the future the investigation of schemes
that treat the response of the LAPW basis with respect to changes
of the potential more efficiently than employing local orbitals.

\begin{acknowledgments}
Financial support from the Deutsche Forschungsgemeinschaft through
the Priority Program 1145 is gratefully acknowledged.

\bibliographystyle{apsrev4-1}
\bibliography{biblio}

\begin{thebibliography}{51}
\expandafter\ifx\csname natexlab\endcsname\relax\def\natexlab#1{#1}\fi
\expandafter\ifx\csname bibnamefont\endcsname\relax
  \def\bibnamefont#1{#1}\fi
\expandafter\ifx\csname bibfnamefont\endcsname\relax
  \def\bibfnamefont#1{#1}\fi
\expandafter\ifx\csname citenamefont\endcsname\relax
  \def\citenamefont#1{#1}\fi
\expandafter\ifx\csname url\endcsname\relax
  \def\url#1{\texttt{#1}}\fi
\expandafter\ifx\csname urlprefix\endcsname\relax\def\urlprefix{URL }\fi
\providecommand{\bibinfo}[2]{#2}
\providecommand{\eprint}[2][]{\url{#2}}

\bibitem[{\citenamefont{Hohenberg and Kohn}(1964)}]{Hohenberg-Kohn}
\bibinfo{author}{\bibfnamefont{P.}~\bibnamefont{Hohenberg}} \bibnamefont{and}
  \bibinfo{author}{\bibfnamefont{W.}~\bibnamefont{Kohn}},
  \bibinfo{journal}{Phys. Rev.} \textbf{\bibinfo{volume}{136}},
  \bibinfo{pages}{B864} (\bibinfo{year}{1964}).

\bibitem[{\citenamefont{{C. Fiolhais, F. Noguiera, and M. A. L.
  Marques}}(2003)}]{DFT-review}
\bibinfo{editor}{\bibnamefont{{C. Fiolhais, F. Noguiera, and M. A. L.
  Marques}}}, ed., \emph{\bibinfo{title}{{A Primer in Density Functional
  Theory}}}, vol. \bibinfo{volume}{620} of \emph{\bibinfo{series}{{Lecutre
  Notes in Physics}}} (\bibinfo{publisher}{Springer, Heidelberg},
  \bibinfo{year}{2003}).

\bibitem[{\citenamefont{{W. Kohn and L. J. Sham}}(1965)}]{Kohn-Sham}
\bibinfo{author}{\bibnamefont{{W. Kohn and L. J. Sham}}},
  \bibinfo{journal}{Phys. Rev.} \textbf{\bibinfo{volume}{140}},
  \bibinfo{pages}{A1133} (\bibinfo{year}{1965}).

\bibitem[{\citenamefont{{D. M. Ceperley and B. J.
  Alder}}(1980)}]{LDA-Ceperley/Adler}
\bibinfo{author}{\bibnamefont{{D. M. Ceperley and B. J. Alder}}},
  \bibinfo{journal}{Phys. Rev. Lett.} \textbf{\bibinfo{volume}{45}},
  \bibinfo{pages}{566} (\bibinfo{year}{1980}).

\bibitem[{\citenamefont{{S. H. Vosko, L. Wilk, and M. Nusair}}(1980)}]{LDA-VWN}
\bibinfo{author}{\bibnamefont{{S. H. Vosko, L. Wilk, and M. Nusair}}},
  \bibinfo{journal}{Can. J. Phys.} \textbf{\bibinfo{volume}{58}},
  \bibinfo{pages}{1200} (\bibinfo{year}{1980}).

\bibitem[{\citenamefont{{J. P. Perdew, K. Burke, and M.
  Ernzerhof}}(1996)}]{GGA-PBE}
\bibinfo{author}{\bibnamefont{{J. P. Perdew, K. Burke, and M. Ernzerhof}}},
  \bibinfo{journal}{Phys. Rev. Lett.} \textbf{\bibinfo{volume}{77}},
  \bibinfo{pages}{3865} (\bibinfo{year}{1996}).

\bibitem[{\citenamefont{{J. P. Perdew and Y. Wang}}(1986)}]{GGA-PW}
\bibinfo{author}{\bibnamefont{{J. P. Perdew and Y. Wang}}},
  \bibinfo{journal}{Phys. Rev. B} \textbf{\bibinfo{volume}{33}},
  \bibinfo{pages}{8800} (\bibinfo{year}{1986}).

\bibitem[{\citenamefont{{J. P. Perdew and M. Levy}}(1983)}]{KS-GAP1}
\bibinfo{author}{\bibnamefont{{J. P. Perdew and M. Levy}}},
  \bibinfo{journal}{Phys. Rev. Lett.} \textbf{\bibinfo{volume}{51}},
  \bibinfo{pages}{1884} (\bibinfo{year}{1983}).

\bibitem[{\citenamefont{{L. J. Sham and M. Schlüter}}(1983)}]{KS-GAP2}
\bibinfo{author}{\bibnamefont{{L. J. Sham and M. Schlüter}}},
  \bibinfo{journal}{Phys. Rev. Lett.} \textbf{\bibinfo{volume}{51}},
  \bibinfo{pages}{1888} (\bibinfo{year}{1983}).

\bibitem[{\citenamefont{{R. W. Godby, M. Schl\"uter, and L. J.
  Sham}}(1988)}]{KS-GAP+DIS3}
\bibinfo{author}{\bibnamefont{{R. W. Godby, M. Schl\"uter, and L. J. Sham}}},
  \bibinfo{journal}{Phys. Rev. B} \textbf{\bibinfo{volume}{37}},
  \bibinfo{pages}{10159} (\bibinfo{year}{1988}).

\bibitem[{\citenamefont{{M. Gr\"{u}ning, A. Marini, and A.
  Rubio}}(2006)}]{KS-GAP+DIS1}
\bibinfo{author}{\bibnamefont{{M. Gr\"{u}ning, A. Marini, and A. Rubio}}},
  \bibinfo{journal}{J. Chem. Phys.} \textbf{\bibinfo{volume}{124}},
  \bibinfo{pages}{154108} (\bibinfo{year}{2006}).

\bibitem[{\citenamefont{K\"ummel and Kronik}(2008)}]{Review:O-functionals}
\bibinfo{author}{\bibfnamefont{S.}~\bibnamefont{K\"ummel}} \bibnamefont{and}
  \bibinfo{author}{\bibfnamefont{L.}~\bibnamefont{Kronik}},
  \bibinfo{journal}{Rev. Mod. Phys.} \textbf{\bibinfo{volume}{80}},
  \bibinfo{pages}{3} (\bibinfo{year}{2008}), \bibinfo{note}{and references
  therein}.

\bibitem[{\citenamefont{{A. G\"orling}}(2005)}]{Review-Goerling}
\bibinfo{author}{\bibnamefont{{A. G\"orling}}}, \bibinfo{journal}{J. Chem.
  Phys.} \textbf{\bibinfo{volume}{123}} (\bibinfo{year}{2005}).

\bibitem[{\citenamefont{G\"orling and Levy}(1994)}]{EXX1}
\bibinfo{author}{\bibfnamefont{A.}~\bibnamefont{G\"orling}} \bibnamefont{and}
  \bibinfo{author}{\bibfnamefont{M.}~\bibnamefont{Levy}},
  \bibinfo{journal}{Phys. Rev. A} \textbf{\bibinfo{volume}{50}},
  \bibinfo{pages}{196} (\bibinfo{year}{1994}).

\bibitem[{\citenamefont{G\"orling and Levy}(1995)}]{EXX2}
\bibinfo{author}{\bibfnamefont{A.}~\bibnamefont{G\"orling}} \bibnamefont{and}
  \bibinfo{author}{\bibfnamefont{M.}~\bibnamefont{Levy}},
  \bibinfo{journal}{Int. J. Quantum Chem.} \textbf{\bibinfo{volume}{56}},
  \bibinfo{pages}{93} (\bibinfo{year}{1995}).

\bibitem[{\citenamefont{G\"orling}(1996)}]{EXX3}
\bibinfo{author}{\bibfnamefont{A.}~\bibnamefont{G\"orling}},
  \bibinfo{journal}{Phys. Rev. B} \textbf{\bibinfo{volume}{53}},
  \bibinfo{pages}{7024} (\bibinfo{year}{1996}), \bibinfo{note}{{Phys.\ Rev.\ B
  \textbf{59}, 10370(E) (1999)}}.

\bibitem[{\citenamefont{{ J. D. Talman and W. F.
  Shadwick}}(1976)}]{OEP-Talman-Shadwick}
\bibinfo{author}{\bibnamefont{{ J. D. Talman and W. F. Shadwick}}},
  \bibinfo{journal}{Phys. Rev. A} \textbf{\bibinfo{volume}{14}},
  \bibinfo{pages}{36} (\bibinfo{year}{1976}).

\bibitem[{\citenamefont{{J. B. Krieger, Y. Li, and G. J.
  Iafrate}}(1992)}]{OEP-atoms}
\bibinfo{author}{\bibnamefont{{J. B. Krieger, Y. Li, and G. J. Iafrate}}},
  \bibinfo{journal}{Phys. Rev. A} \textbf{\bibinfo{volume}{45}},
  \bibinfo{pages}{101} (\bibinfo{year}{1992}).

\bibitem[{\citenamefont{{Y. Li, J. B. Krieger, and G. J.
  Iafrate}}(1993)}]{OEP-atoms-1}
\bibinfo{author}{\bibnamefont{{Y. Li, J. B. Krieger, and G. J. Iafrate}}},
  \bibinfo{journal}{Phys. Rev. A} \textbf{\bibinfo{volume}{47}},
  \bibinfo{pages}{165} (\bibinfo{year}{1993}).

\bibitem[{\citenamefont{{T. Kotani}}(1994)}]{EXX-LMTO-Kotani-I}
\bibinfo{author}{\bibnamefont{{T. Kotani}}}, \bibinfo{journal}{Phys. Rev. B}
  \textbf{\bibinfo{volume}{50}}, \bibinfo{pages}{14816} (\bibinfo{year}{1994}).

\bibitem[{\citenamefont{{ T. Kotani}}(1995)}]{EXX-LMTO-Kotani-II}
\bibinfo{author}{\bibnamefont{{ T. Kotani}}}, \bibinfo{journal}{Phys. Rev.
  Lett.} \textbf{\bibinfo{volume}{74}}, \bibinfo{pages}{2989}
  (\bibinfo{year}{1995}).

\bibitem[{\citenamefont{{M. Städele, J. A. Majewski, P. Vogl, and A.
  Görling}}(1997)}]{EXX-PP-Staedele-1}
\bibinfo{author}{\bibnamefont{{M. Städele, J. A. Majewski, P. Vogl, and A.
  Görling}}}, \bibinfo{journal}{Phys. Rev. Lett.}
  \textbf{\bibinfo{volume}{79}}, \bibinfo{pages}{2089} (\bibinfo{year}{1997}).

\bibitem[{\citenamefont{{M. Städele, M. Moukara, J. A. Majewski, P. Vogl, and
  A. Görling}}(1999)}]{EXX-PP-Staedele-2}
\bibinfo{author}{\bibnamefont{{M. Städele, M. Moukara, J. A. Majewski, P. Vogl,
  and A. Görling}}}, \bibinfo{journal}{Phys. Rev. B}
  \textbf{\bibinfo{volume}{59}}, \bibinfo{pages}{10031} (\bibinfo{year}{1999}).

\bibitem[{\citenamefont{{ A. Fleszar}}(2001)}]{EXX-PP-Fleszar}
\bibinfo{author}{\bibnamefont{{ A. Fleszar}}}, \bibinfo{journal}{Phys. Rev. B}
  \textbf{\bibinfo{volume}{64}}, \bibinfo{pages}{245204}
  (\bibinfo{year}{2001}).

\bibitem[{\citenamefont{{S. Sharma, J. K. Dewhurst, and C.
  Ambrosch-Draxl}}(2005)}]{Sharma}
\bibinfo{author}{\bibnamefont{{S. Sharma, J. K. Dewhurst, and C.
  Ambrosch-Draxl}}}, \bibinfo{journal}{Phys. Rev. Lett.}
  \textbf{\bibinfo{volume}{95}}, \bibinfo{pages}{136402}
  (\bibinfo{year}{2005}).

\bibitem[{\citenamefont{{E. Engel}}(2009)}]{Engel}
\bibinfo{author}{\bibnamefont{{E. Engel}}}, \bibinfo{journal}{Phys. Rev. B}
  \textbf{\bibinfo{volume}{80}}, \bibinfo{pages}{161205(R)}
  (\bibinfo{year}{2009}).

\bibitem[{\citenamefont{{A. Makmal, R. Armineto, E. Engel, L. Kronik, and S.
  Kümmel}}(2009)}]{Makmal}
\bibinfo{author}{\bibnamefont{{A. Makmal, R. Armineto, E. Engel, L. Kronik, and
  S. Kümmel}}}, \bibinfo{journal}{Phys. Rev. B} \textbf{\bibinfo{volume}{80}},
  \bibinfo{pages}{161204(R)} (\bibinfo{year}{2009}).

\bibitem[{\citenamefont{{A. Hesselmann, A. W. G\"otz, F. D. Sala, and A.
  G\"orling}}(2007)}]{Balance-Gaussian}
\bibinfo{author}{\bibnamefont{{A. Hesselmann, A. W. G\"otz, F. D. Sala, and A.
  G\"orling}}}, \bibinfo{journal}{J. Chem. Phys.}
  \textbf{\bibinfo{volume}{127}}, \bibinfo{pages}{054102}
  (\bibinfo{year}{2007}).

\bibitem[{\citenamefont{{A. G\"orling, A. Hesselmann, M. Jones, and M.
  Levy}}(2008)}]{Hesselmann}
\bibinfo{author}{\bibnamefont{{A. G\"orling, A. Hesselmann, M. Jones, and M.
  Levy}}}, \bibinfo{journal}{J. Chem. Phys.} \textbf{\bibinfo{volume}{128}},
  \bibinfo{pages}{104104} (\bibinfo{year}{2008}).

\bibitem[{\citenamefont{{A. Hesselmann and A.
  G\"orling}}(2008)}]{Comparison-OEP-KS}
\bibinfo{author}{\bibnamefont{{A. Hesselmann and A. G\"orling}}},
  \bibinfo{journal}{Chem. Phys. Lett} \textbf{\bibinfo{volume}{455}},
  \bibinfo{pages}{110} (\bibinfo{year}{2008}).

\bibitem[{Fle()}]{Fleur}
\bibinfo{howpublished}{\url{http://www.flapw.de}}.

\bibitem[{\citenamefont{{R. T. Sharp and G. K. Horton}}(1953)}]{OEP/OPM}
\bibinfo{author}{\bibnamefont{{R. T. Sharp and G. K. Horton}}},
  \bibinfo{journal}{Phys. Rev.} \textbf{\bibinfo{volume}{90}},
  \bibinfo{pages}{317} (\bibinfo{year}{1953}).

\bibitem[{\citenamefont{{V. Sahni, J. Gruenebaum, and J. P.
  Perdew}}(1982)}]{OEP<=>EXX}
\bibinfo{author}{\bibnamefont{{V. Sahni, J. Gruenebaum, and J. P. Perdew}}},
  \bibinfo{journal}{Phys. Rev. B} \textbf{\bibinfo{volume}{26}},
  \bibinfo{pages}{4371} (\bibinfo{year}{1982}).

\bibitem[{\citenamefont{{E. Wimmer, H. Krakauer, M. Weinert, and A. J.
  Freeman}}(1981)}]{FLAPW1}
\bibinfo{author}{\bibnamefont{{E. Wimmer, H. Krakauer, M. Weinert, and A. J.
  Freeman}}}, \bibinfo{journal}{Phys. Rev. B} \textbf{\bibinfo{volume}{24}},
  \bibinfo{pages}{864} (\bibinfo{year}{1981}).

\bibitem[{\citenamefont{{M. Weinert, E. Wimmer, and A. J.
  Freeman}}(1982)}]{FLAPW2}
\bibinfo{author}{\bibnamefont{{M. Weinert, E. Wimmer, and A. J. Freeman}}},
  \bibinfo{journal}{Phys. Rev. B} \textbf{\bibinfo{volume}{26}},
  \bibinfo{pages}{4571} (\bibinfo{year}{1982}).

\bibitem[{\citenamefont{{H. J. F. Jansen and A. J. Freeman}}(1984)}]{FLAPW3}
\bibinfo{author}{\bibnamefont{{H. J. F. Jansen and A. J. Freeman}}},
  \bibinfo{journal}{Phys. Rev. B} \textbf{\bibinfo{volume}{30}},
  \bibinfo{pages}{561} (\bibinfo{year}{1984}).

\bibitem[{\citenamefont{{D. Singh}}(1991)}]{Local_Orbitals1}
\bibinfo{author}{\bibnamefont{{D. Singh}}}, \bibinfo{journal}{Phys. Rev. B}
  \textbf{\bibinfo{volume}{43}}, \bibinfo{pages}{6388} (\bibinfo{year}{1991}).

\bibitem[{\citenamefont{{E. E. Krasovskii, A. N. Yaresko, and V. N.
  Antonov}}(1994)}]{Local_Orbitals2}
\bibinfo{author}{\bibnamefont{{E. E. Krasovskii, A. N. Yaresko, and V. N.
  Antonov}}}, \bibinfo{journal}{J. Electron Spectrosc. Relat. Phenom.}
  \textbf{\bibinfo{volume}{68}}, \bibinfo{pages}{157} (\bibinfo{year}{1994}).

\bibitem[{\citenamefont{{C. Friedrich, A. Schindlmayr, S. Bl\"ugel, and T.
  Kotani}}(2006)}]{Local_Orbitals3}
\bibinfo{author}{\bibnamefont{{C. Friedrich, A. Schindlmayr, S. Bl\"ugel, and
  T. Kotani}}}, \bibinfo{journal}{Phys. Rev. B} \textbf{\bibinfo{volume}{74}},
  \bibinfo{pages}{045104} (\bibinfo{year}{2006}).

\bibitem[{\citenamefont{Andersen}(1975)}]{Local_Orbitals_Andersen}
\bibinfo{author}{\bibfnamefont{O.~K.} \bibnamefont{Andersen}},
  \bibinfo{journal}{Phys. Rev. B} \textbf{\bibinfo{volume}{12}},
  \bibinfo{pages}{3060} (\bibinfo{year}{1975}).

\bibitem[{\citenamefont{{T. Kotani and M. van
  Schilfgaarde}}(2002)}]{Kotani-MixedBasis}
\bibinfo{author}{\bibnamefont{{T. Kotani and M. van Schilfgaarde}}},
  \bibinfo{journal}{Solid State Commun.} \textbf{\bibinfo{volume}{121}},
  \bibinfo{pages}{461} (\bibinfo{year}{2002}).

\bibitem[{\citenamefont{{M. Betzinger, C. Friedrich, and S.
  Blügel}}(2010)}]{PBE0-NonLocalExactExchangePotential}
\bibinfo{author}{\bibnamefont{{M. Betzinger, C. Friedrich, and S. Blügel}}},
  \bibinfo{journal}{Phys. Rev. B} \textbf{\bibinfo{volume}{19}},
  \bibinfo{pages}{195117} (\bibinfo{year}{2010}).

\bibitem[{\citenamefont{{C. Friedrich, S. Bl\"ugel, and A.
  Schindlmayr}}(2010)}]{GW-MixedBasis}
\bibinfo{author}{\bibnamefont{{C. Friedrich, S. Bl\"ugel, and A.
  Schindlmayr}}}, \bibinfo{journal}{Phys. Rev. B}
  \textbf{\bibinfo{volume}{81}}, \bibinfo{pages}{125102}
  (\bibinfo{year}{2010}).

\bibitem[{\citenamefont{{C. Friedrich, A. Schindlmayr, and S.
  Blügel}}(2009)}]{CoulombMatrix-MixedBasis}
\bibinfo{author}{\bibnamefont{{C. Friedrich, A. Schindlmayr, and S. Blügel}}},
  \bibinfo{journal}{{Comput. Phys. Comm.}} \textbf{\bibinfo{volume}{180}},
  \bibinfo{pages}{347} (\bibinfo{year}{2009}).

\bibitem[{\citenamefont{{E. Engel, S. Keller, A. Facco Bonetti, H. M\"uller,
  and R. M. Dreizler}}(1995)}]{RELKS1}
\bibinfo{author}{\bibnamefont{{E. Engel, S. Keller, A. Facco Bonetti, H.
  M\"uller, and R. M. Dreizler}}}, \bibinfo{journal}{Phys. Rev. A}
  \textbf{\bibinfo{volume}{52}}, \bibinfo{pages}{2750} (\bibinfo{year}{1995}).

\bibitem[{\citenamefont{{E. Engel, S. Keller, and R. M.
  Dreizler}}(1996)}]{RELKS2}
\bibinfo{author}{\bibnamefont{{E. Engel, S. Keller, and R. M. Dreizler}}},
  \bibinfo{journal}{Phys. Rev. A} \textbf{\bibinfo{volume}{53}},
  \bibinfo{pages}{1367} (\bibinfo{year}{1996}).

\bibitem[{\citenamefont{{E. Engel, A. Facco Bonetti, S. Keller, I.
  Andrejkovics, and R. M. Dreizler}}(1998)}]{RELKS3}
\bibinfo{author}{\bibnamefont{{E. Engel, A. Facco Bonetti, S. Keller, I.
  Andrejkovics, and R. M. Dreizler}}}, \bibinfo{journal}{Phys. Rev. A}
  \textbf{\bibinfo{volume}{58}}, \bibinfo{pages}{964} (\bibinfo{year}{1998}).

\bibitem[{\citenamefont{{A. Hesselmann}}()}]{Hesselmann-privat}
\bibinfo{author}{\bibnamefont{{A. Hesselmann}}}, \bibinfo{note}{private
  communication, Troullier-Martins pseudopotentials constructed from the atomic
  EXX potential are employed}.

\bibitem[{\citenamefont{{R. J. Magyar, A. Fleszar, and E. K. U.
  Gross}}(2004)}]{EXX-Ar/Ne}
\bibinfo{author}{\bibnamefont{{R. J. Magyar, A. Fleszar, and E. K. U. Gross}}},
  \bibinfo{journal}{Phys. Rev. B} \textbf{\bibinfo{volume}{69}},
  \bibinfo{pages}{045111} (\bibinfo{year}{2004}).

\bibitem[{\citenamefont{{T. C. Chiang and F. J.
  Himpsel}}(1989)}]{Landolt-Boernstein}
\bibinfo{author}{\bibnamefont{{T. C. Chiang and F. J. Himpsel}}},
  \emph{\bibinfo{title}{{Band structure and core levels of tetrahedrally-bonded
  semiconductors}}}, vol. \bibinfo{volume}{23 a} of
  \emph{\bibinfo{series}{{Landolt-B\"ornstein - Group III Condensed Matter
  Numerical Data and Functional Relationships in Science and Technology}}}
  (\bibinfo{publisher}{Springer Verlag, Berlin}, \bibinfo{year}{1989}).

\bibitem[{\citenamefont{{M. Runne and G. Zimmerer}}(1995)}]{Ne-Ar}
\bibinfo{author}{\bibnamefont{{M. Runne and G. Zimmerer}}},
  \bibinfo{journal}{{Nuclear Instruments and Methods in Physics Research
  Section B: Beam Interactions with Materials and Atoms}}
  \textbf{\bibinfo{volume}{101}}, \bibinfo{pages}{156} (\bibinfo{year}{1995}).

\end{thebibliography}

\end{acknowledgments}

\end{document}